\documentclass[11pt,a4paper]{article}
\usepackage{jcappub}
\usepackage{amsmath}
\usepackage{amsfonts}
\usepackage{amssymb}
\usepackage{graphicx}
\usepackage{caption}
\usepackage{subcaption}
\usepackage{textcomp}
\usepackage{geometry}
\usepackage{color}
\usepackage{slashbox}
\usepackage{multirow}
\graphicspath{{.},{plots/}}

\def\alt{\raise0.3ex\hbox{$\;<$\kern-0.75em\raise-1.1ex\hbox{$\sim\;$}}}
\def\agt{\raise0.3ex\hbox{$\;>$\kern-0.75em\raise-1.1ex\hbox{$\sim\;$}}}

\title{The sensitivity of Cherenkov telescopes to dark matter and
  astrophysical anisotropies in the diffuse gamma-ray
  background}

\author[a,b]{Joachim Ripken,}
\author[a,c,d]{Alessandro Cuoco,}
\author[c,d,e]{Hannes-S. Zechlin,}
\author[a,f]{Jan Conrad}
\author[e]{Dieter Horns}

\affiliation[a]{The Oskar Klein Centre for Cosmo Particle Physics, \\ AlbaNova,
 SE-106 91 Stockholm, Sweden}
\affiliation[b]{Max-Planck-Institute for Solar System Research,\\ 
  Max-Planck-Stra\ss e 2, D-37191 Katlenburg-Lindau, Germany}
\affiliation[c]{Department of Physics, University of Torino, via P. Giuria 1, 10125 Torino, Italy}
\affiliation[d]{Istituto Nazionale di Fisica Nucleare, via P. Giuria 1, 10125 Torino, Italy}
\affiliation[e]{University of Hamburg, Institut f\"ur Experimentalphysik,\\
Luruper Chaussee 149, D-22761 Hamburg, Germany}
\affiliation[f]{Wallenberg Academy Fellow}
\emailAdd{ripken@mps.mpg.de}
\emailAdd{cuoco@fysik.su.se}
\emailAdd{zechlin@to.infn.it}
\emailAdd{conrad@fysik.su.se}
\emailAdd{dieter.horns@physik.uni-hamburg.de}

\arxivnumber{1211.6922}

\abstract{In this article, the capability of present (H.E.S.S., MAGIC,
  VERITAS) and planned (CTA) ground-based Cherenkov telescope systems
  for detecting angular anisotropies in the diffuse gamma-ray
  background is investigated. Following up on a study of the impact of
  instrumental characteristics (effective area, field of view, angular
  resolution, and background rejection efficiency), the first part
  examines the influence of different observational strategies, i.e.
  whether a single deep observation or a splitting over multiple
  shallow fields is preferred. In the second part, the sensitivity to
  anisotropies generated by self-annihilating dark matter is studied
  for different common dark matter models. We find that a relative
  contribution of $\sim\!10\%$ from dark matter annihilation to the
  extra-galactic diffuse gamma-ray background can be detected with
  planned configurations of CTA. In terms of the thermally-averaged
  self-annihilation cross section, the sensitivity of CTA corresponds
  to values below the thermal freeze-out expectation $\langle \sigma v
  \rangle = 3 \times 10^{-26} \, \text{cm}^{3} \text{s}^{-1}$ for dark
  matter particles lighter than $\sim$200\,GeV. We stress the
  importance of constraining anisotropies from unresolved
  astrophysical sources with currently operating instruments already,
  as a novel and complementary method for investigating the properties
  of TeV sources.}

\begin{document}
\maketitle
\flushbottom

\section{Introduction}
The study of gamma-ray anisotropies
\cite{Ackermann:2012uf,Cuoco:2012yf} has recently provided new and
complementary insights into the nature of gamma-ray sources and the
extra-galactic diffuse gamma-ray background (EDGB)
\cite{Abdo:2010nz,Dermer:2007fg}. Experimentally, EDGB means the
residual large-scale isotropic emission measured at high galactic
latitudes after subtracting the \emph{galactic} diffuse emission.
This emission arises mostly from the integrated contribution of
unresolved extra-galactic sources and, possibly, from the annihilation
or decay of dark matter (DM).  A contribution from galactic sources is
however also possible if their emission extends to sufficiently high
galactic latitudes to produce an almost isotropic contribution.  It
has been argued, for example, that millisecond pulsars can give a
contribution to the EDGB \cite{FaucherGiguere:2009df}. Similarly, also
galactic DM could contribute to the EDGB, besides the extra-galactic
one. The measured energy spectrum of the EDGB is, however, compatible
with a simple featureless power law \cite{Abdo:2010nz} so that
complementary information, for example from anisotropy, can help
isolating different contributions to this emission. The pattern of
anisotropies has been studied with different techniques, mainly
through its angular power spectrum (APS), as in
\cite{Ackermann:2012uf,Cuoco:2012yf}. Likewise, the study of the
1-point probability distribution function (PDF) \cite{Malyshev:2011zi}
and the cross-correlation with galaxy catalogues \cite{Xia:2011ax}
provide complementary information.

It has been argued that dark matter self-annihilation or decay
could leave a specific imprint on the anisotropy pattern and spectrum
of the EDGB
\cite{Ando:2005xg,Ando:2006cr,Cuoco:2006tr,Cuoco:2007sh,SiegalGaskins:2008ge,Fornasa:2009qh,SiegalGaskins:2009ux,Ando:2009fp,Zavala:2009zr,Hensley:2009gh,Cuoco:2010jb,Fornasa:2012gu}.
In fact, while the emissivity of ordinary astrophysical sources scales
with the inner source densities $\sim\!\varrho$ (modulo a source-class
dependent bias factor), the emissivity of self-annihilating DM scales
with its density squared $\sim\!\varrho^{2}$. Owing to this
difference, self-annihilating DM could leave its signature in the
angular power spectrum of the EDGB. This simple picture can be further
complicated by the presence of unresolved point sources that produce a
Poissonian-like APS, more closely resembling the DM one. Nonetheless,
even if the astrophysical Poissonian term dominates the intrinsic
clustering APS, revealing the astrophysical and dark matter APS to be
similar, it is still possible to separate the two contributions by
measuring their energy dependence, i.e. the \emph{anisotropy energy
  spectrum} \cite{SiegalGaskins:2009ux}. The different hypotheses
about the origin of the EDGB can thus be tested by measuring both the
angular power spectrum and its energy dependence.

In addition, theoretical predictions for the anisotropies generated by
different gamma-ray source populations (e.g., blazars and galaxy
clusters~\cite{clusters}, millisecond
pulsars~\cite{SiegalGaskins:2010mp}, star-forming
galaxies~\cite{Ando:2009nk}) become available, extending our knowledge
on this approach and its potential.

Gamma-rays are currently detected mainly with two
techniques. Observatories in space, such as the Fermi Large Area
Telescope (Fermi-LAT) \cite{Atwood:2009ez}, enable the detection of
gamma-rays through pair conversion in the detector itself. With the
Fermi-LAT, gamma-ray photons can be observed in the energy range from
a few ten MeV up to a few hundred GeV, with an effective area close to
$1\,\text{m}^{2}$ and a field of view (fov) of $\sim\!  1 \,
\text{sr}$.  Fermi-LAT routinely operates in sky surveying mode,
continuously mapping the entire sky within $\sim$3\,h. Complementary,
Cherenkov light emitted from air showers initiated by gamma-rays
penetrating the upper atmosphere can be observed with ground-based
telescopes, such as H.E.S.S. \cite{Aharonian:2006pe}, MAGIC
\cite{Albert:2007xh}, and VERITAS \cite{Aliu:2011zi}, or the planned
Cherenkov telescope array (CTA) \cite{CTA:2010,Doro:2012xx}. The
effective collection area of ground-based instruments is typically of
the order of $10^5$\,m$^2$ to $10^6$\,m$^2$. The energy range of
current experiments lies between $60 \, \text{GeV}$ and $100 \,
\text{TeV}$, but future realizations of this concept will lower the
threshold to $10 \, \text{GeV}$ or even $5 \, \text{GeV}$
\cite{Aharonian:2000rf}. Contrary to Fermi-LAT, ground-based
instruments only offer a relatively small fov of typically a few msr,
so that all-sky scans are not feasible.  Observations can only be
pursued during darkness under the condition of a clear sky, reducing
the duty-cycle to $\sim$1\,000\,h per year. Furthermore, the trigger
rate is dominated by a large background of hadronic showers.  Specific
techniques are employed to reduce this background substantially.
However, even after sufficient gamma-hadron separation, air showers
induced by cosmic-ray electrons can still contribute significantly to
the background at a few hundred GeV
\cite{Abdo:2009zk,Ackermann:2010ij,Aharonian:2008aa,
  Aharonian:2009ah}, hardly separable from photon-induced air showers.
Despite the difficulties mentioned above, interesting scales for the
investigation of anisotropies are typically very small (less than
$1^\circ$), so that a small fov does not pose a serious limit for
their study. The angular resolution of Cherenkov telescopes is
typically better than $0.1^{\circ}$, i.e. multipoles in the range
between 100 and 1\,000 can be easily resolved. At the same time, the
background is expected to be isotropic at small scales and therefore
no fundamental obstacle either.

In this paper, we investigate the capabilities of ground-based
Cherenkov telescopes for measuring gamma-ray angular anisotropies.  In
section \ref{sec_toymc}, we introduce a simplified Monte-Carlo
approach that is used to study the impact of instrumental
characteristics (fov, angular resolution, and background rejection
power) to the detection sensitivity for anisotropies. Given that a
combined analysis of data collected from observations of several
different targets would be most effective, we investigate the
influence of the \textit{observational strategy}, i.e. whether it
would be advantageous or disadvantageous to split the data into
several fov. In section \ref{bench}, we list the instrumental setups
that are used as benchmarks, and we provide estimates of the CR
background rates expected for these configurations. Section
\ref{sec_DM} introduces a more realistic simulation setup and a
sensitivity study to anisotropies from DM self-annihilation for
different DM models. Final comments and discussion are provided in
section \ref{sec_dis}.

\section{Optimizing the observational strategy} \label{sec_toymc} 

\subsection{Simplified setup}
As a measure of anisotropy we use the angular power spectrum (APS) of
fluctuations throughout this paper.  Given a map $I(\vartheta,
\varphi)$ on the sphere, the fluctuation map is defined as
$\delta(\vartheta, \varphi) = I(\vartheta, \varphi)/ \bar{I} - 1$,
where $\bar{I}$ denotes the mean value of $I$.  Thus, by definition,
the mean value of $\delta(\vartheta, \varphi) $ is $0$.  The
fluctuation map $\delta(\vartheta, \varphi)$ is decomposed into
spherical harmonics $Y^{\ell}_m(\vartheta, \varphi)$ as
$\delta(\vartheta, \varphi) = \sum_{\ell m}a_{\ell m}
Y^{\ell}_m(\vartheta, \varphi)$, where $a_{\ell m}$ denote the
coefficients of the spherical harmonic decomposition, $\ell =
0,\dots,\infty$, and $m = -l, \dots, l$. The coefficients $a_{\ell m}$
define the APS by
\begin{equation} C_\ell \equiv \langle|a_{\ell
    m}|^2\rangle,
\end{equation}
where $\langle\dots\rangle$ indicates the statistical ensemble
average. Then, the quantity
\begin{equation}\label{pspectrum}
  \hat{C_\ell}= \sum_m \frac{|a_{\ell m}|^2}{2 l+1}
\end{equation}
provides an unbiased estimator of the true power spectrum $C_\ell$, i.e.
$\langle\hat{C_\ell}\rangle=C_\ell$.

Note that apart from the dimensionless fluctuation APS the
dimension-full APS of the map $I(\vartheta,\varphi)$ itself can be
used. This is particularly useful for the analysis of real data sets
(see \cite{Ackermann:2012uf}), but we refrain from further
consideration.

In order to simulate event lists containing anisotropies, we generate
sky maps with a given $C_{\ell}$ spectrum.  $C_{\ell}$ can be
interpreted as the width of the $a_{\ell m}$ distribution over $m$ for
a fixed $\ell$.  Assuming Gaussian fluctuations, the $a_{\ell m}$
coefficients for a fixed $\ell$ can be randomly chosen from a Gaussian
distribution centered on $0$ with a width $\sqrt{C_{\ell}}$. The phase
is chosen equally distributed between $0$ an $2 \pi$ with the
condition $a_{\ell m} = a_{-\ell m}^{\star}$, to ensure fluctuation
maps with real values. Twelve realizations of the $a_{\ell m}$ are
generated for a given spectrum, and thus twelve independent
fluctuation maps $\delta(\vartheta, \varphi)$.  Moreover, we simulate
maps with five different benchmark APS following a power law with
slopes $s = 0.5$, $1.0$, $1.5$, $2.0$, $2.5$, i.e.  $\ell (\ell+1)
C_{\ell} \sim \ell^s$. The angular resolution of the simulated sky
maps is chosen to be $0.002^{\circ}$, corresponding to a maximum
resolvable multipole $\ell = 9 \times 10^{4}$. The maps are normalized
as $\delta^\prime(\vartheta, \varphi)= (\delta(\vartheta,
\varphi)-\delta_{\rm min})/(\delta_{\rm max}-\delta_{\rm min})$, in
order to obtain a distribution between 0 and 1 to simulate
events. Hence, the map $\delta^\prime$ acts as effective intensity map
$I'(\vartheta, \varphi) \equiv \delta^\prime(\vartheta, \varphi)$. We
emphasize that the anisotropies of this effective intensity map are
independent of the original $\delta(\vartheta, \varphi)$ normalization
and always in the range $-1<\delta I'/\bar{I'}<1$ by definition, thus
implying large fluctuations of the order of $100\%$. The setup has
been optimized for the purpose of investigating instrumental effects
on the APS. However, the setup is extended to a more realistic
approach in section \ref{sec_DM}, allowing the choice of an arbitrary
anisotropy level.

Based upon the template intensity maps $I'(\vartheta, \varphi)$, the
simulation of events requires the definition of the following three
parameters: $\sigma_{\text{fov}}$, the half-width of the camera
acceptance, $\sigma_{\text{psf}}$, the width of the point-spread
function (PSF), and the signal fraction $f_\text{sig}$, the ratio of
signal events with respect to the sum of signal and background
events. Event lists contain an anisotropic \textit{signal} component
and a \textit{background} component. The latter is isotropic by
definition. Both the camera acceptance and the PSF are assumed to
follow Gaussian distributions.

Each event of a list is handled in three subsequent steps: The
celestial position is chosen randomly according to the camera
acceptance. Comparing a uniform deviate with $f_\text{sig}$, it is
then decided whether the event is treated as a \textit{signal} or a
\textit{background} event. If the event belongs to the background, the
event is just retained. In the signal case, instead, a variate $z$ is
generated, following a normalized uniform distribution. If $z$ is
smaller than $I'(\vartheta, \varphi)$ at the event's position, the
event is kept, otherwise it is rejected. Eventually, each event is
randomly displaced from its original direction according to the PSF,
in order to realize the convolution of the map with the PSF. If not
particularly specified, an event list contains $N_\text{ev} = 10^{7}$
entries. We point out that this number is unrealistically high, even
for CTA, and is used to isolate and emphasize instrumental effects
only. Simulations with a more realistic number of signal and
background events are provided in section \ref{sec_DM}.

We use the HEALPix software and pixelization scheme
\cite{Gorski:2004by} to create and analyze count maps (with
$N_{\text{pix}}$ pixels) as well as to extract the APS. Before the
analysis, a count map is cast into a fluctuation map
\begin{equation}
  \delta''(\vartheta,\varphi) = 
  \frac{N_{\text{pix}}}{N_\text{ev}} \left[ \sum_{i =
    1}^{N_{\text{pix}}} x_{i} \, b_{i}(\vartheta,\varphi)
    \right] - 1,
\end{equation}
where $x_{i}$ denotes the number of events in pixel $i$, and
$b_{i}(\vartheta,\varphi)$ equals $1$ inside pixel $i$ and $0$
otherwise. In this way, $\delta''(\vartheta,\varphi)$ is normalized
such that $\int \mathrm{d} \Omega \ [1\!+\!
  \delta''(\vartheta,\varphi)] = 4 \pi$, where $\mathrm{d}\Omega$ is
the differential solid angle element in spherical coordinates. It
should be noted that, strictly speaking, $\delta''(\vartheta,\varphi)$
represents a fluctuation map
(i.e. $\langle\delta''(\vartheta,\varphi)\rangle=0$) only in the case
that the average is performed over the whole sky rather than over the
region of interest where the counts are located. The normalization has
been chosen to keep a simple form of the noise term in the recovered
APS (see next section), i.e. $C_N=4\pi/N_\text{ev}$, where no factors
of $f_\text{sky}=\Omega_\text{fov}/4\pi$ are involved
($\Omega_\text{fov}$ denotes the solid angle of the fov). We remark
that a direct analysis of a fluctuation map built from a raw count map
is pursued, while in principle the fluctuation map built from the
reconstructed flux map (the count map divided by the exposure of the
experiment or the fov in our case) can be analyzed. While the latter
is preferred for real data analyses, the former is sufficiently
accurate for a sensitivity study. It implies the extracted APS to be a
convolution of the experiment's windowing function (the APS of the
camera acceptance) with the true signal. For the large $\ell$
considered below, the windowing effect is marginal.

The statistical uncertainty of the APS is derived from the twelve
simulations that have been performed for each case. In the following
plots, the RMS as estimates of the standard deviation are shown as
uncertainty bands.

\begin{figure}[t]
\centering
    \begin{subfigure}[t]{0.47\textwidth}
    \centering
    \includegraphics[width=\textwidth]{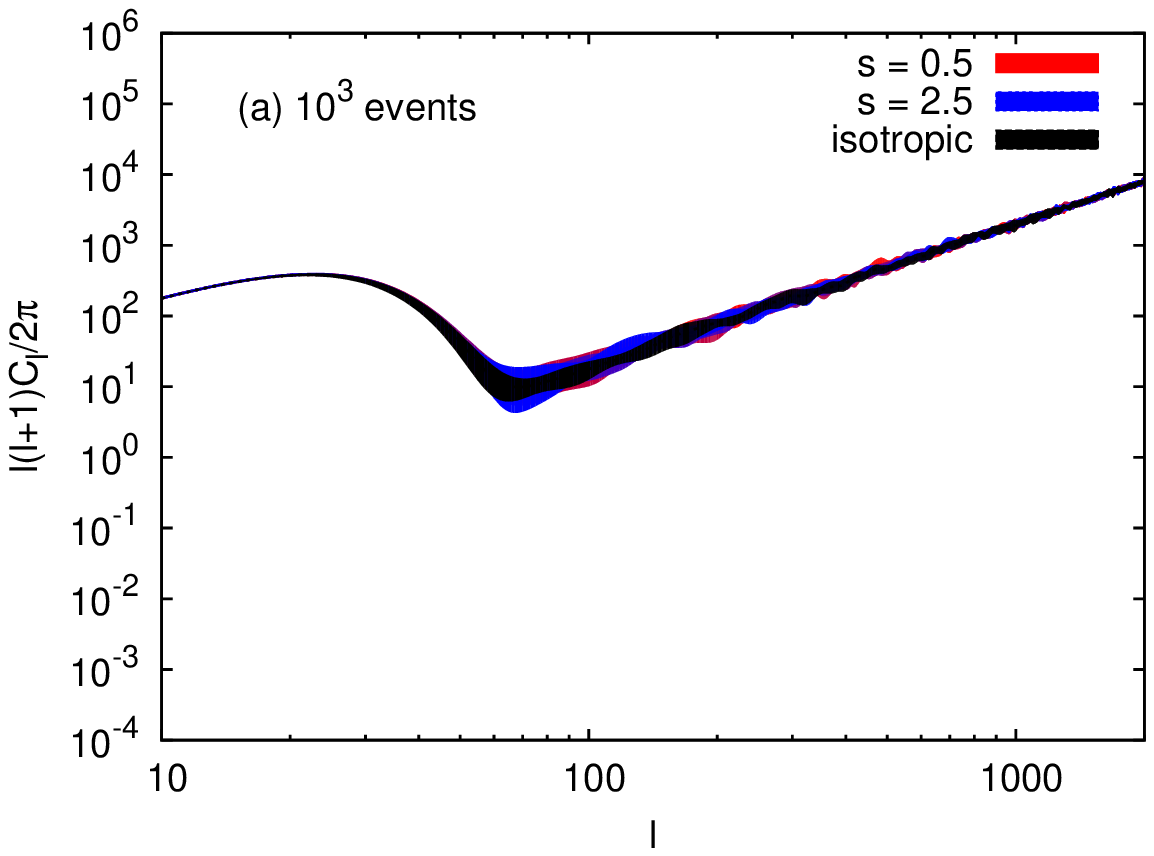}
    \caption{}
    \end{subfigure}
    \hspace{0.3cm}
    \begin{subfigure}[t]{0.47\textwidth}
    \centering
    \includegraphics[width=\textwidth]{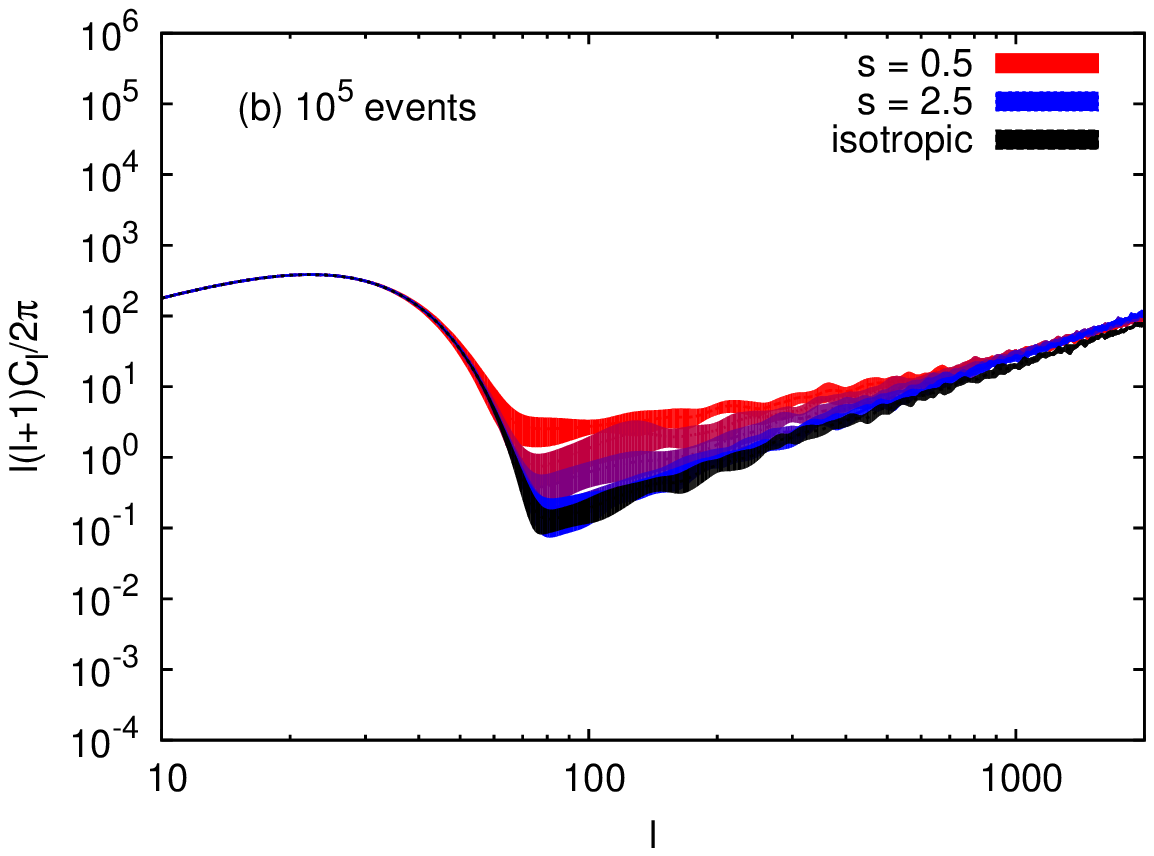}
    \caption{}
    \end{subfigure}
    \begin{subfigure}[t]{0.47\textwidth}
    \centering
    \includegraphics[width=\textwidth]{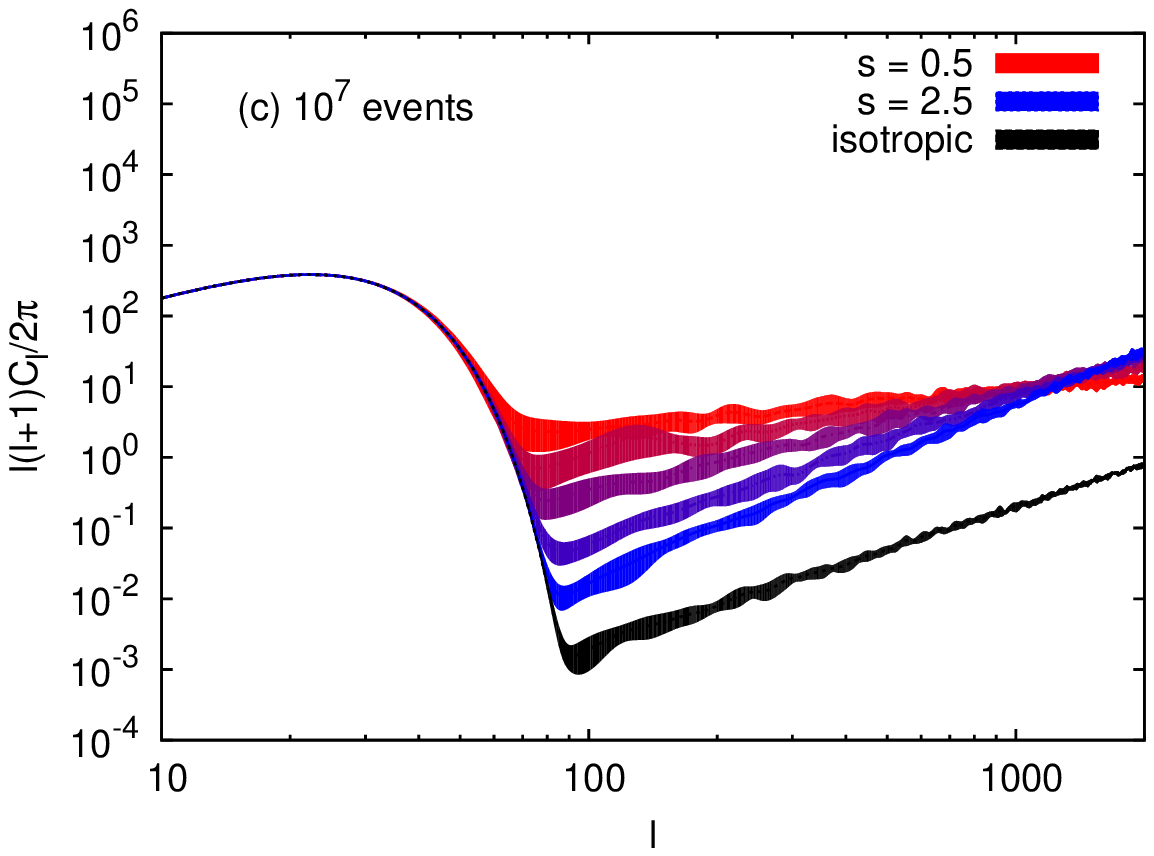}
    \caption{}
    \end{subfigure}
    \hspace{0.3cm}
    \begin{subfigure}[t]{0.47\textwidth}
    \centering
    \includegraphics[width=\textwidth]{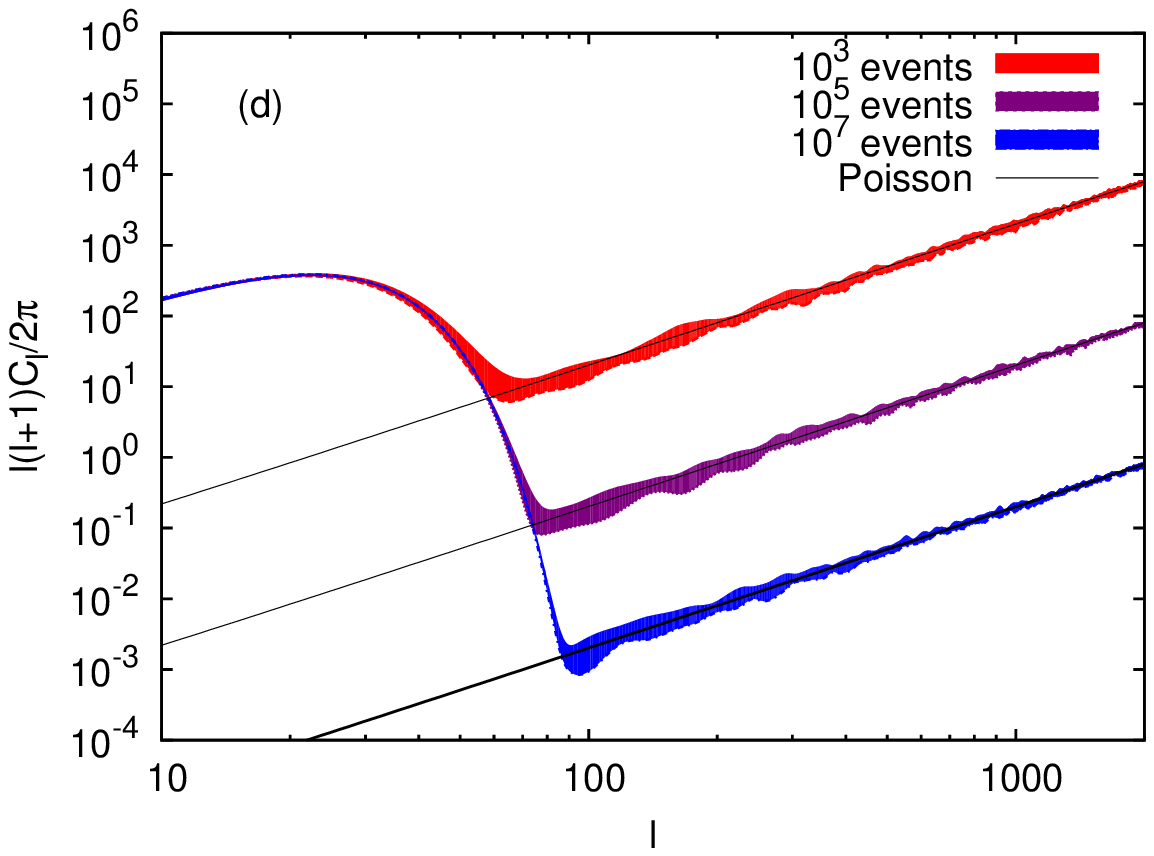}
    \caption{}
    \end{subfigure}
  \caption{Angular power spectra measured from simulated event maps
    with different input APS slopes $s$ between $0.5$ and $2.5$ in
    steps of $\Delta s = 0.5$ (from red to blue).  The simulations
    contain signal-only events.  The lower, black APS, labeled
    ``isotropic'', refers to a background-only simulation with the
    same number of events.  In this figure, PSF effects are neglected
    and $\sigma_{\text{fov}} = 2.5^{\circ}$.  Sub-figures represent
    the cases of (a) $10^3$ events, (b) $10^5$ events, and (c) $10^7$
    events. The effect of an increasing event number on the Poissonian
    noise is depicted in (d), which shows together the APS of the
    background-only simulations from the panels (a), (b), and (c).}
  \label{fig_eventnumber}
\end{figure}

\subsection{Influence of the detector configuration} \label{sec_res}
In figure~\ref{fig_eventnumber}, examples of APS measured from the
simulated maps with given input-APS slopes are shown for a varying
number of events. No background events have been included and all
events are of signal type, except for the lowest, black APS, which is
measured from a simulation containing background events only (an
isotropic simulation).  For small $\ell$, the windowing function
dominates the spectrum, which is distorted accordingly. For larger
$\ell$ we see that the measured APS is the sum of the Poissonian noise
and of the intrinsic APS. If the number of events is large enough, the
different signal slopes can be easily discriminated.  If the number of
events is small, however, the random noise from the finite number of
events prevails the measured APS. The Poissonian noise is given by
$C_N = 4 \pi /N_\text{ev}$ for full sky coverage (see appendix B in
\cite{Cuoco:2006tr} and section IV.A in \cite{Ando:2005xg}), where
$N_\text{ev}=N_\mathrm{sig}\! +\!  N_\mathrm{bg}$ is the total number
of events. This is in good agreement with the noise estimates from the
background-only simulations (and partial sky-coverage corrected APS,
see previous section), as shown in the lower right panel of
figure~\ref{fig_eventnumber}. Since the Poissonian noise term is
known, it is customary to remove it from the estimated APS so that to
better show the intrinsic APS. We will discuss noise-subtracted APS
for realistic simulations in section \ref{sec_DM}. The same
considerations apply to the PSF effect and the background fraction
discussed in the following, which can all be modeled, and thus
corrected to give a final, unfolded APS.  In the forward-folding
approach used in section \ref{sec_DM}, these corrections are in
principle not required, since the model is directly convolved with the
PSF before comparison with the simulated data. Nonetheless, for
illustration purposes, we show totally unfolded APS in
section~\ref{sec_DM}.

\begin{figure}[t]
\centering
    \begin{subfigure}[t]{0.47\textwidth}
    \centering
    \includegraphics[width=\textwidth]{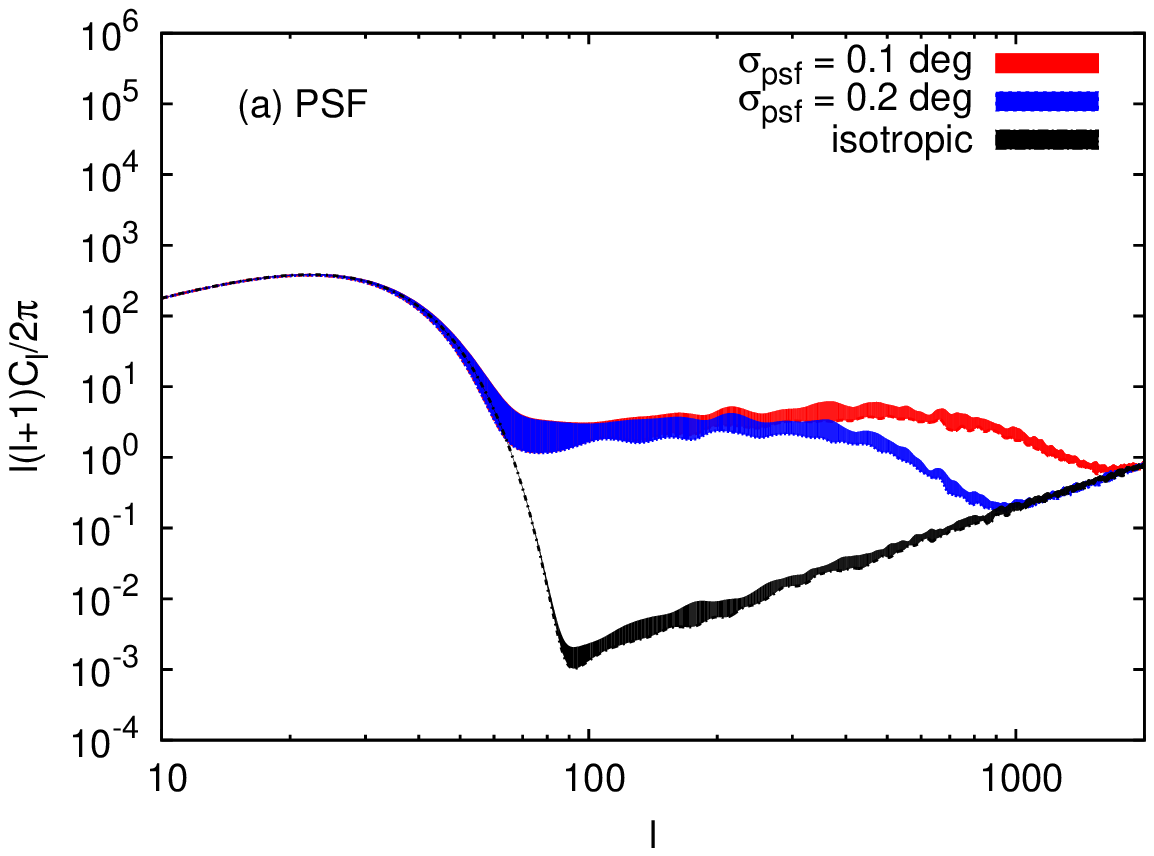}
    \caption{}
    \end{subfigure}
    \hspace{0.3cm}
    \begin{subfigure}[t]{0.47\textwidth}
    \centering
    \includegraphics[width=\textwidth]{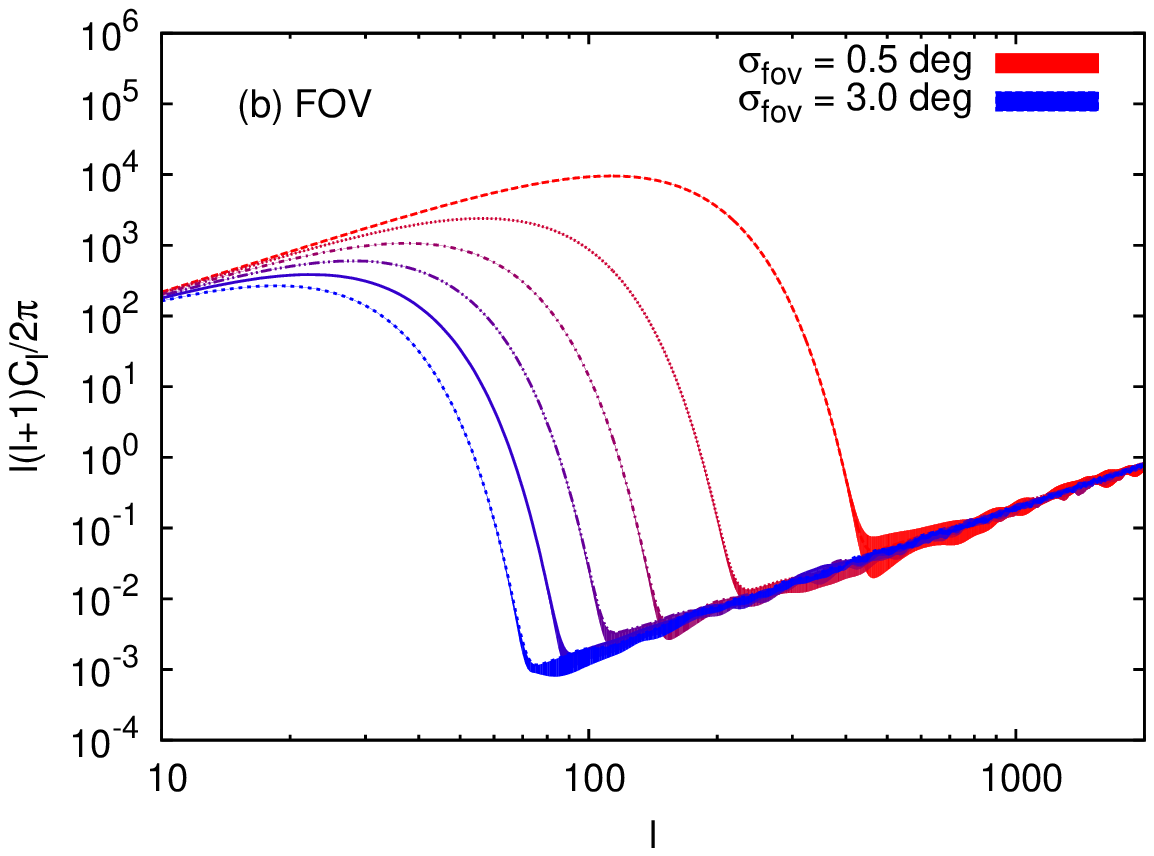}
    \caption{}
    \end{subfigure}
    \begin{subfigure}[t]{0.47\textwidth}
    \centering
    \includegraphics[width=\textwidth]{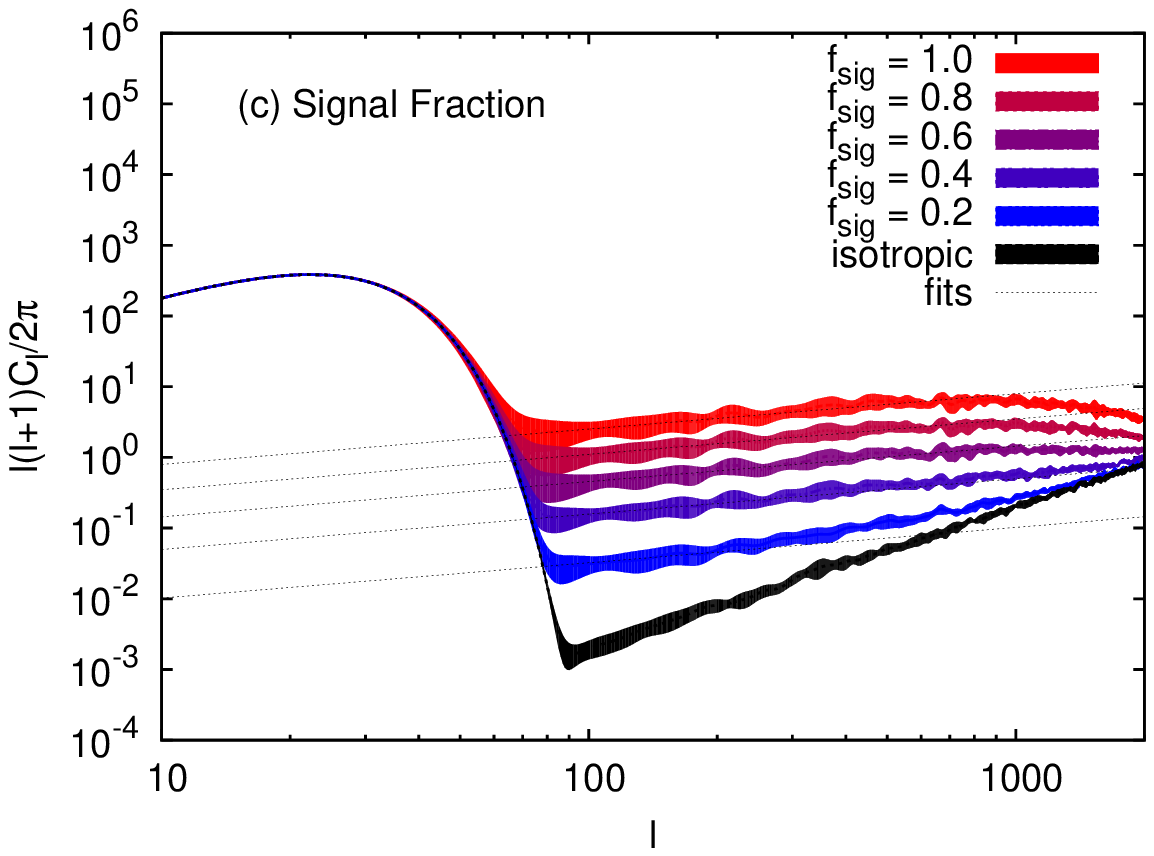}
    \caption{}
    \end{subfigure}
  \caption{(a): influence of the PSF on the measured APS. The slope of
    the input spectrum is $s=0.5$.  The chosen PSF widths are
    $\sigma_\text{psf}=0.1^{\circ}$ (red band) and
    $\sigma_\text{psf}=0.2^{\circ}$ (blue band). (b): influence of the
    fov on the APS for a pure background (isotropic) event list, with
    $\sigma_{\text{psf}} = 0.05^{\circ}$. The width
    $\sigma_\mathrm{fov}$ is increased in steps of 0.5, in the range
    from $\sigma_\mathrm{fov}=0.5^\circ$ (red) to
    $\sigma_\mathrm{fov}=3.0^\circ$ (blue). (c): influence of the
    signal fraction $f_\text{sig}$ on the measured APS for an input
    slope of $s=0.5$; background events are distributed isotropically.
    Dotted lines show the APS in case of vanishing PSF, vanishing fov
    distortions, and vanishing noise.}
  \label{fig_collage}
\end{figure}

The effect of the instrument's PSF on the measured APS is illustrated
in the top-left panel of figure~\ref{fig_collage}: The PSF suppresses
the signal at large $\ell$ and drives it toward the level of the
Poissonian noise. The characteristic multipole of the downturn is
related to the PSF width $\sigma_{\text{psf}}$ by $\ell_{s} \approx
180^{\circ}/\sigma_{\text{psf}}$. This effect can be corrected if
$\sigma_{\text{psf}}$ (or the full PSF shape in a non-Gaussian case)
is known \cite{Ackermann:2012uf}, however at the expense of an
increasing uncertainty on the measured APS.

Anisotropies at an angular scale larger than the fov are
suppressed. This effect is illustrated in the top-right panel of
figure~\ref{fig_collage}, where the APS of an isotropic
(background-only) event list is shown for different fov. A larger fov
allows exploring larger scales and thus lower multipoles. The minimum
resolvable multipole is approximately given by $\ell_{\text{min}}
\approx 180^{\circ}/\sigma_{\text{fov}}$.

The bottom panel of figure~\ref{fig_collage} illustrates the effect of
different signal fractions $f_\text{sig}$. The background component
mainly originates from two different processes:
\begin{enumerate}
\item[(i)] Events caused by mostly isotropic cosmic rays (protons and
  electrons), which have been misclassified as photons. Anisotropies
  in the hadronic cosmic-ray background are indeed present at the
  level of $10^{-4}$
  \cite{Abbasi:2011ai,Abbasi:2011zka,Abdo:2008kr,Abdo:2008aw,Vernetto:2009xm}. However,
  they extend to a multipole of $\sim\!20$ only \cite{Abbasi:2011ai}
  and are thus negligible in our analysis. Anisotropies in the
  electron background have not been detected so far
  \cite{Ackermann:2010ip,Aguilar:2013qda}. For Cherenkov telescope
  systems the hadronic cosmic-ray background component dominantes the
  gamma-ray signal. Therefore, a sufficiently good gamma-hadron
  separation is a crucial characteristic of the instrument. To reduce
  the residual background of cosmic-ray electrons, which becomes
  important at energies below a few hundred GeV, a sufficiently good
  gamma-electron separation would be favorable as well. However,
  gamma-electron separation capabilities are limited due to the
  similarity of electron and photon initiated showers. With current
  instruments a rejection of $\sim\!50\%$ of the electrons seems
  possible \cite{Aharonian:2008aa, Aharonian:2009ah}, while the
  expected performance of CTA on this aspect has not been studied in
  detail yet.
\item[(ii)] An intrinsically isotropic component of the diffuse photon
  background, which does not act as signal according to our
  definition. For instance, cosmological photons produced by truly
  diffuse processes may account for this component.
\end{enumerate}

Note that we do not consider the effect of the local variations of the
\emph{night-sky background} and its effect on the camera acceptance
\cite{Rowell:2003jb}. Rather, we assume that prominent features in the
fov such as bright stars can be eventually masked and excluded from
the analysis. A more realistic MC simulation would be required to
study this effect in detail. This is left for future work.

The bottom panel of figure~\ref{fig_collage} shows the measured APS
for different signal factions. A slope of $s=0.5$ has been assumed for
the input spectrum used to simulate the signal part of the maps and
event lists. Dotted lines show the measured APS in the ideal case of
vanishing PSF effects, vanishing fov distortion, and vanishing noise
(but with background still included). As expected, the background
fraction has a large impact and considerably reduces the signal height
with respect to the intrinsic Poissonian noise, i.e. the
signal-to-noise ratio. For the shown case of $10^7$ events a signal
fraction of 20\% is still easily detectable. However, the
signal-to-noise ratio decreases rapidly with decreasing $f_{\rm sig}$,
and for realistic cases $f_{\rm sig}$ can be as low as 0.1\% while
reaching a value of 2\% in optimistic scenarios (see table 2 in
section 4). We thus conclude that a good \textit{background rejection}
is crucial.

\begin{figure}[t]
\centering
    \begin{subfigure}[t]{0.455\textwidth}
    \centering
    \includegraphics[width=\textwidth]{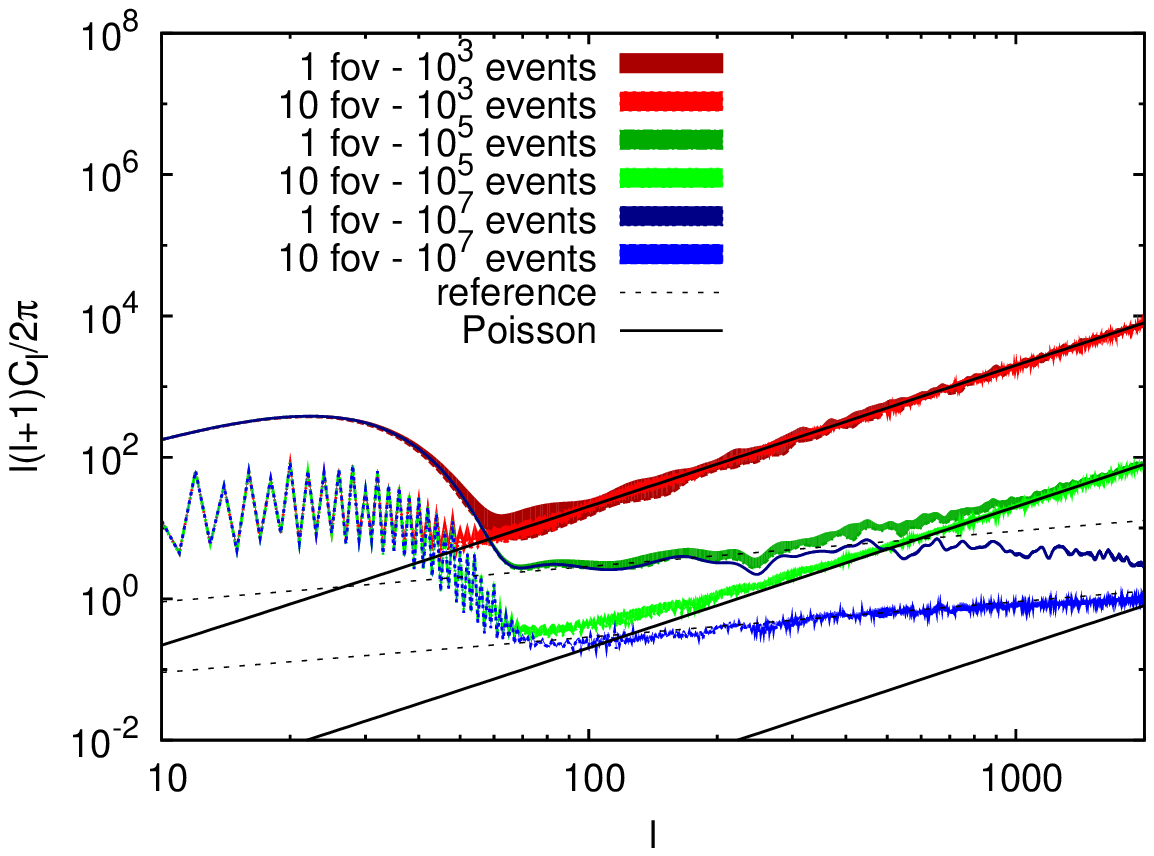}
    \caption{}
    \end{subfigure}
    \hspace{0.3cm}
    \begin{subfigure}[t]{0.47\textwidth}
    \centering
    \includegraphics[width=\textwidth]{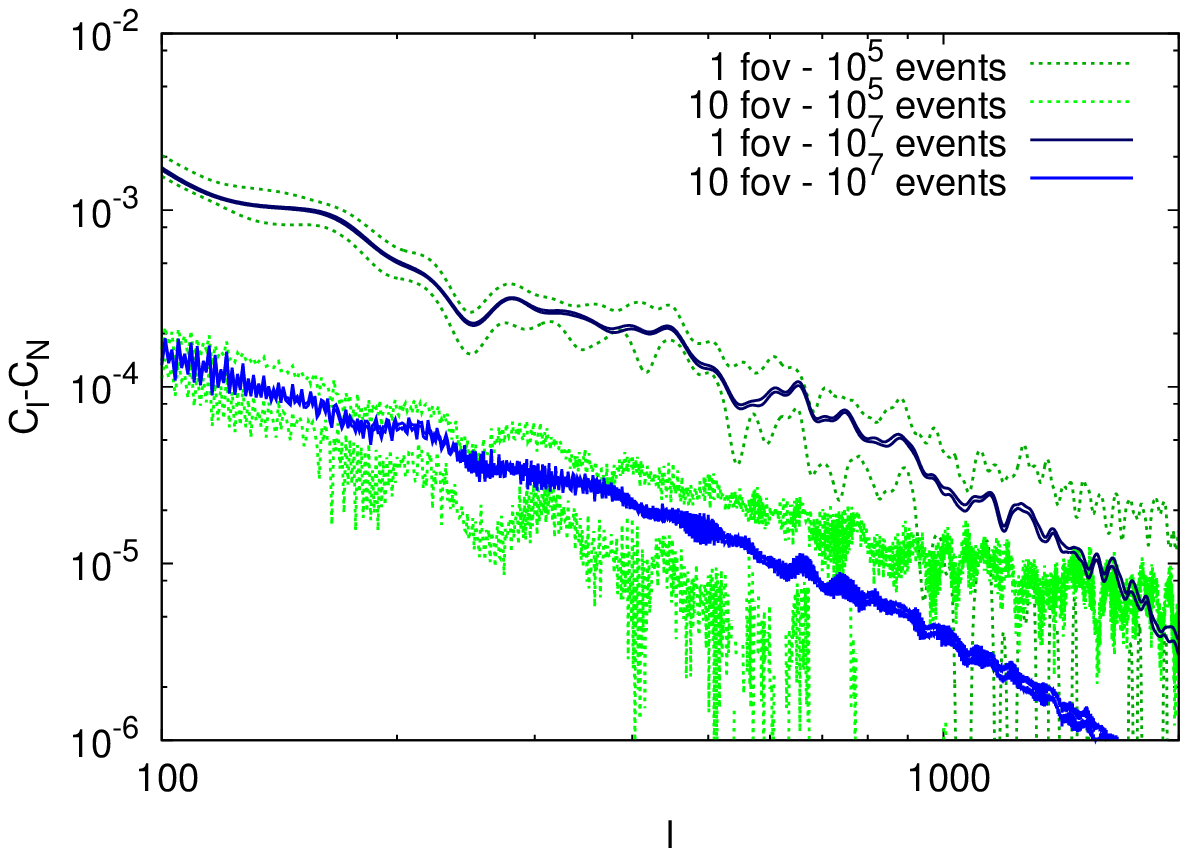}
    \caption{}
    \end{subfigure}
  \caption{Impact of the observational strategy. (a): the measured APS
    for an input spectrum with a slope $s=0.5$ is shown for $10^{3}$
    (red), $10^{5}$ (green), and $10^{7}$ (blue) signal-only events,
    distributed in a single fov (dark colors) or in ten different fov
    (light colors), respectively. Note that with our definition of the
    normalization of the maps, the Poissonian noise (depicted by the
    solid black lines) depends only on the number of events and it is
    the same in the single or multiple fov cases. For comparison, the
    thin dotted lines show the input APS for the two $10^{7}$-event
    cases. A Gaussian PSF with a width of 0.05$^\circ$ is
    assumed. (b): same as above, but showing the noise-subtracted
    power $C_{\ell}-C_N$ on the $y$-axis instead of
    $\ell(\ell+1)C_\ell/2\pi$. The green (blue) lines enclose the
    1$\sigma$ error regions on the measured APS for $10^5$ ($10^7$)
    events, distributed in a single (dark colors) or in ten different
    fov (light colors).}
  \label{fig_strategy}
\end{figure}

Finally, we investigate the effect of the observational strategy. In
general, the detection of anisotropies requires a large amount of
observation time. Thus, dedicated observations could eventually not be
feasible, given different targets competing for the limited
observation time available. The use of combined observations obtained
on different targets would thus be preferable. This approach is
investigated in figure~\ref{fig_strategy}. Here, a number of $10^{3}$,
$10^{5}$, and $10^7$ signal-only events are distributed within one fov
as well as ten different ones.  Qualitatively, when the same number of
events is distributed in a single or multiple fov two competing
effects arise: On the one hand, the density of signal events decreases
since they are spread over a larger area, and thus the signal-to-noise
ratio decreases (with our definition of the normalization of the maps,
the Poissonian noise depends only on the number of events and it is
the same in the single or multiple fov cases). On the other hand,
multiple fov result in an increased number of modes available for the
APS calculation, reducing the statistical error on the measured APS
(on the sphere this effect is known as \emph{cosmic variance}) and
thus improving the sensitivity. The overall number of events
eventually determines the dominating error and thus the dominating
effect.

As illustrated in figure~\ref{fig_strategy}, multiple fov indeed
decrease the error on the measured APS and reduce the signal-to-noise
ratio. The $10^7$-events case shows that a highly significant
detection over all multipoles is achieved both in the single and
multiple fov cases.  On the other hand, the $10^5$-events case shows
that while with the single fov observation a non zero APS can be
detected up to a multipole of $\sim$1000, with the multiple fov case
it is possible to significantly detect the anisotropy only up to a
multipole of $\sim$ 400.  In fact, while the error is decreased, the
signal-to-noise ratio becomes too low to get a significant detection.
Finally, in the $10^3$-events case there is no significant detection
both in the single or multiple fov.  A quantitative numerical study of
this effect in a realistic scenario including background is presented
in the section \ref{sec_DM}.  Analytical formulae are derived in
appendix A.

It should be stressed that the use of multiple fov requires the
systematic error in the absolute calibration of each fov to be kept
reasonably under control. In the following, we assume that this
systematic error can be neglected compared to the statistical ones. A
better assessment of this uncertainty will likely be available from
future performance studies of CTA.

\section{Benchmark instrumental setups and cosmic-ray backgrounds}\label{bench}
Simulations of two instrument classes are presented in the
following. The used instrumental setups are motivated by the
characteristics of currently operating instruments such as H.E.S.S.,
MAGIC, and VERITAS, and the expected properties of CTA. Two different
threshold energies (100\,GeV and 300\,GeV) have been adopted,
revealing different signal-to-background ratios.

\begin{table}[t]
\centering
\begin{tabular}{|lcccccc|}
\hline
\multirow{2}{*}{IACT} & \multicolumn{3}{c}{$E_\mathrm{thr}=100$\,GeV} & \multicolumn{3}{c|}{$E_\mathrm{thr}=300$\,GeV} \\
  & $A_\mathrm{eff}$ [m$^2$] & $\sigma_\mathrm{fov}$ [deg] & $\sigma_\mathrm{psf}$ [deg] & $A_\mathrm{eff}$ [m$^2$] & $\sigma_\mathrm{fov}$ [deg] & $\sigma_\mathrm{psf}$ [deg] \\
\hline
current & \multicolumn{3}{c}{---} & $10^5$ & $1.7$ & $0.1$ \\
CTA & $10^5$ & $3,\,4,\,5$ & $0.05$ & $3\times 10^5$ & $3,\,4,\,5$ & $0.05$ \\
\hline
\end{tabular}
\caption{Characteristics (effective area $A_\mathrm{eff}$, field of
  view $\sigma_\mathrm{fov}$, and PSF $\sigma_\mathrm{psf}$) of the
  benchmark instrumental setups used in the simulations. The
  setups have been chosen in accordance with characteristics of
  current-generation IACTs and the planned CTA
  observatory. Predictions are made for two different threshold
  energies $E_\mathrm{thr}$, 100\,GeV and 300\,GeV, respectively.}
\label{tab_bench}
\end{table} 

For currently operating instruments, the performance below 1\,TeV
typically degrades rapidly in energy. Hence, a threshold energy of
300\,GeV is considered only. Above 300 GeV, an effective area of
$10^5$\,m$^2$ (after selection cuts which improve the fraction of
gamma rays with respect to hadrons) is assumed. So is a fov of
1.7$^\circ$ and an angular resolution of 0.1$^\circ$
\cite{hesspoland,Aharonian:2006pe,Albert:2007xh,Aliu:2011zi}.

For CTA, recent Monte-Carlo studies of the performance
\cite{CTA:2010,Bernlohr:2012we,Vandenbroucke:2011sh} indicate an
effective area of $3 \times 10^5$\,m$^2$ above 300\,GeV and
$10^5$\,m$^2$ above 100\,GeV (see figure 15 in
\cite{Bernlohr:2012we}). In both cases, we assume an angular
resolution of 0.05$^\circ$ (see figures 10 and 17 in
\cite{Bernlohr:2012we}). The effective area as well as the angular
resolution improve with energy, but the simulations have been
considerably simplified choosing constant values close to the
threshold energies, representing a conservative choice. For the fov,
values between between 3$^\circ$ and 5$^\circ$ are considered (see
table 3 in \cite{Bernlohr:2012we}). We emphasize that a rather large
fov of $\mathcal{O}(5^\circ)$ might be provided by the types C, D, and
J of the suggested CTA arrays \cite{Bernlohr:2012we}. In addition, an
effectively larger fov can be achieved with dedicated pointing
patterns, adjusting the pointings of individual telescopes to
correspondingly different offsets from a targeted position (see figure
3c in \cite{CTA:2010}). Table~\ref{tab_bench} summarizes the
characteristics of the instruments used in this study. The performance
of the recently built H.E.S.S.-II array is expected to lie in between
the two setups considered here \cite{hesspoland}.

The intensity of the isotropic hadronic background component depends
on analysis cuts and the quality of the gamma-hadron separation. With
respect to present instruments, CTA will provide improvement in the
performance of the hadron rejection process. However, a substantial
part of the low-energy background is made by cosmic-ray electrons,
which are more difficult to separate from gamma rays. The study in
\cite{Bernlohr:2012we} provides a simulation of the total expected
background from hadronic and leptonic components, which are used to
estimate the background corresponding to our setups. In particular, we
refer to their figure 16 of the integrated background rate per beam
above a given energy threshold (and for the assumed effective area).

For a threshold energy of 100\,GeV, a background rate of 0.01\,Hz to
0.03\,Hz per beam yields a total background rate of 50\,Hz to 150\,Hz
for an angular resolution of $\sim\!0.1^\circ$ and for a fov of
5$^\circ$. Thus, benchmark background rates of 10\,Hz and 100\,Hz are
assumed in this study, the first being slightly optimistic (but still
possible depending on eventual improvements of the background
rejection), and the second represents a more conservative
choice. Rates for different fov scale with the factor
$(\sigma_\mathrm{fov,1}/\sigma_\mathrm{fov,2})^2$. For simplicity, we
use 10\,Hz and 100\,Hz as benchmark background rates for all
considered fov, although these rates are over-estimated for smaller
fov of 3$^\circ$ and 4$^\circ$. The background for a threshold of
300\,GeV covers a range between $3 \times 10^{-4}$\,Hz and $3 \times
10^{-3}$\,Hz per beam, and the angular resolution covers values
between 0.06$^\circ$ and 0.1$^\circ$. For a fov of 5$^\circ$, this
corresponds to a total background rate of 1.5\,Hz up to 42\,Hz. Again,
an optimistic and a more conservative background rate are chosen,
i.e., 1\,Hz and 10\,Hz, respectively.

Assuming the same characteristics as for CTA for threshold energies
above 300 GeV for current IACTs, the scaling to the correspondingly
smaller fov reduces the background rate by a factor of 10. A further
reduction in the rate is given by the smaller effective area. However,
the background rejection capabilities are inferior with respect to the
expectation for CTA, thus increasing the background rate. We use the
same benchmark background rates as for CTA with threshold energies
above 300 GeV, i.e. 1\,Hz and 10\,Hz. These values match typically
observed background rates.

\section{Dark matter sensitivity}\label{sec_DM}
A more realistic setup can now be employed to simulate maps with a
given level of anisotropy. Here, we consider an anisotropy spectrum
with a slope of $s=2$ only, i.e. the same slope as of Poissonian
noise.  With the conventions given in section~\ref{sec_toymc}, $s=2$
corresponds to an APS constant in multipole and can therefore be
characterized by a single number, i.e. $C_{\ell}=C_P$ for all
$\ell$. This kind of anisotropy spectrum, known as Poisson anisotropy,
is typically expected from unresolved point sources and provides a
good approximation for most of the common DM scenarios
\cite{Ando:2005xg,Ando:2006cr,Cuoco:2007sh,SiegalGaskins:2008ge,Fornasa:2009qh,SiegalGaskins:2009ux,Ando:2009fp,Zavala:2009zr,Hensley:2009gh,Cuoco:2010jb,Fornasa:2012gu}. The
similarity to the Poissonian noise also suggests a straightforward way
to simulate this kind of anisotropy: For $N$ identical sources
distributed all over the sky, the Poisson anisotropy of the
fluctuation map will be $C_P=4\pi/N$. Inverting the process, a map
with an anisotropy equal to $C_P$ can be simulated distributing
$N=4\pi/C_P$ equal sources on the sphere. This method is adopted in
the following. In general, unresolved sources are not identical but
have a certain flux distribution $\mathrm{d}N/\mathrm{d}S$, which
typically follows a power law or broken power-law
distribution. Anisotropy measurements can be used to recover the
underlying $\mathrm{d}N/\mathrm{d}S$
\cite{Cuoco:2012yf,Malyshev:2011zi}.  To check our approximation of
using an effective delta-like $\mathrm{d}N/\mathrm{d}S$, simulations
for a realistic $\mathrm{d}N/\mathrm{d}S$ have been performed, and are
described later in this section. We follow the algorithm described in
section~\ref{sec_toymc} to produce a sequence of background or signal
events from the simulated source maps. Different to
section~\ref{sec_toymc}, the source maps are not rescaled, since they
possess an intrinsic anisotropy normalization that we want to retain.

The total number of background events is given by the integrated
background rates estimated in the previous section. The gamma-ray flux
is normalized to the EDGB measured with Fermi-LAT: $\phi(E)=\phi_0\,
(E/100\, \text{MeV})^{-2.41}$, with $\phi_0 = 1.45 \times
10^{-7}$\,cm$^{-2}$\,s$^{-1}$\,sr$^{-1}$\,MeV$^{-1}$
\cite{Abdo:2010nz}. Extrapolating the power-law spectrum above the
considered threshold energies of 100\,GeV and 300\,GeV gives the
integral fluxes $\phi(E > 100 \, \text{GeV}) \, \approx \, 6 \times
10^{-10}$\,cm$^{-2}$\,s$^{-1}$\,sr$^{-1}$ $ \,$ and $ \,\;$ $\phi(E >
300 \, \text{GeV}) \approx 1.3 \times
10^{-10}$\,cm$^{-2}$\,s$^{-1}$\,sr$^{-1}$. The effect of the EDGB
attenuation expected from pair production on the extragalactic
background light is neglected here. The expected softening is mild in
most of the attenuation models
\cite{Dominguez:2010bv,Finke:2009xi,Franceschini:2008tp,Gilmore:2011ks,Kneiske:2010pt,Stecker:2012ta,Meyer:2012us},
and taking it into account would only slightly reduce the total flux
above 100\,GeV, while the attenuation could be more pronounced above
300\,GeV. For a CTA observation of 100\,h, this results in $10\,448$
and $6\,659$ signal events in total, assuming a fov of 5$^\circ$ and a
threshold energy of 100\,GeV and 300\,GeV, respectively (setups as
discussed in table~\ref{tab_bench}).  The events are distributed
between DM and astrophysical sources according to the relative
contribution to the EDGB as considered in the following. Changing the
threshold energy from 100\,GeV to 300\,GeV, the number of gamma-ray
events is reduced by less than a factor of 2, while the number of
background events is reduced by a factor of 10 (due to their steeper
energy spectrum and improved background rejection at higher
energies). The signal-to-noise ratio of a given data set can thus be
improved considering a threshold energy of 300\,GeV. The numbers of
gamma rays and background events for all the setups are reported in
table~\ref{tab_sens}.


The intrinsic anisotropy of astrophysical sources is modeled in
accordance to the recent measurement of anisotropy performed with
Fermi-LAT in the range of 1\,GeV to 50\,GeV \cite{Ackermann:2012uf}.
The fluctuation energy spectrum of the measured anisotropy is
compatible with a constant value of $\sim 10^{-5}$, while the
\emph{intensity} energy spectrum of anisotropy is compatible with a
power law with slope $\sim2.4$. In combination, these results indicate
that the measured anisotropy originates from unresolved blazars. This
is further supported by the analysis performed in
\cite{Cuoco:2012yf}. We assume that the result holds above
100\,GeV. Thus, a value of $C_P^A=10^{-5}$ is used for the intrinsic
astrophysical source anisotropies. However, also other values of
$C_P^A$ will be explored to assess the robustness of the results with
respect to the choice of this parameter.

The theoretical predictions for the intrinsic DM anisotropy are
uncertain and span over a few orders of magnitude ranging from
$10^{-4}$ to $10^{-1}$. We assume a benchmark value $C_P^{DM}=10^{-3}$
and will comment on other values in the following. A simplified
analytic calculation is reported in the appendices A, B and C,
illustrating the expected dependence of the sensitivity on the choice
of $C_P^{DM}$ and $C_P^{A}$.

To estimate the sensitivity to the DM component, we vary the relative
contribution of DM to the total EDGB flux, assigning the remaining
flux to the astrophysical component. In particular, we simulate 20
values of the DM contribution $p$ from $0 \%$ to $100 \%$ in steps of
$5 \%$.  The average astrophysical spectrum $C_{\ell}^{\text{A}}$, its
error $\sigma_{C_{\ell}}$, and the average spectrum
$C_\ell^{\text{A}+\text{DM},p\%}$ composed of both the astrophysical
and a fractional DM contribution of $p$ are estimated from $20$
realizations each. Note that $C_{\ell}^{\text{A}}$ indicates the
\emph{measured} average APS as function of the multipole $\ell$.
Ideally, assuming that no biases are present in the simulation
pipeline, and after correcting for the instrumental effects, the CR
background, and the Poisson noise, we should observe that $\langle
C_{\ell}^{\text{A}} \rangle = C_P^{A}$ for all $\ell$, where
$\langle\dots\rangle$ indicates the average over many simulations.
The same consideration applies to $C_\ell^{\text{A}+\text{DM},p\%}$.

To quantify the sensitivity to the relative DM contribution, we use a
simple chi-square approach, comparing two different definitions:
\begin{equation}\label{eq:chi1}
  \chi^{2}_i(p) = \sum_{\ell = 100}^{1000}
  \left(\frac{C_{\ell}^{\text{A}+\text{DM},p\%}-C_{\ell}^{\text{A},i}}
  {\sigma_{C_\ell}}\right)^2,
\end{equation}
\begin{equation}\label{eq:chi2}
  \chi^{2}_j(p) = \sum_{\ell = 100}^{1000} 
  \left(\frac{C_{\ell}^{\text{A}+\text{DM},p\%,j}-C_{\ell}^\text{A}}
  {\sigma_{C_\ell}}\right)^2,
\end{equation}
where $C_{\ell}^{\text{A}+\text{DM},p\%,j}$ and
$C_{\ell}^{\text{A},i}$ are the spectra from the single realizations
as opposed to the average ones, $C_\ell^{\text{A}+\text{DM},p\%}$ and
$C_\ell^\text{A}$, respectively. Only multipoles above 100 are used,
discarding lower multipoles affected by the fov of the instrument.
The quantity $\chi^{2}(p)$ follows a chi-square distribution with
$901$ degrees of freedom, so that we can quote sensitivities at $95\%$
confidence level (CL) for the value of $p$ corresponding to a
$\chi^{2}(p)$ larger than 972.  Given that $\chi^{2}(p)$ scatters
among different realizations, an additional criterion must be
specified. For example, the average value of $\chi_{i}^{2}(p)$ over
$i$ can be used.  Here, we adopt the more conservative requirement
that at least 19 out of the 20 different $\chi_i^{2}(p)$ values (for a
given $p$) are larger than the sensitivity threshold of 972, in order
to set the value of $p$ corresponding to the $95 \%$ CL.
The two different ways to define $\chi^{2}(p)$, see
Eqs. \eqref{eq:chi1} and \eqref{eq:chi2}, correspond to two different
definitions of the sensitivity. In the first approach, we assume and
simulate a ``true'' case without a DM component and search for the
minimum (at $95 \%$ CL) DM fraction required to exclude the null
hypothesis. In the second approach, we assume and simulate ``true''
cases including a DM contribution of $p$ and search for the minimum
contribution for which the null hypothesis becomes incompatible with
the simulated data. Both cases are found to give consistent results.
We point out that the definition of sensitivity used here is in short
a ``$95 \%$ CL incompatibility with the null hypothesis'', which
implies a comparison of two $\chi^{2}$-distributions. Another commonly
employed definition of a sensitivity as ``$95 \%$ upper limit in the
case of a null detection outcome of the experiment'' (which requires
interval estimation through a profile likelihood or test statistic
procedure) can give more stringent sensitivities, but is not
considered in this work.

\begin{table}[t]
\begin{center}
\footnotesize{
 \textbf{H.E.S.S./MAGIC/VERITAS \; E$_{\rm \bf th}$=300 GeV  \;  ${\bf \sigma_\mathrm{\textbf{fov}}=1.7^\circ}$ }  
\begin{tabular}{|llrrc|}
  \hline
  \multirow{2}{*}{Observation time [h]} & \multirow{2}{*}{Bg. rate [Hz]} & 
  \multirow{2}{*}{Sens.} & \multirow{2}{*}{$N_\mathrm{sig}$} & \multirow{2}{*}{$N_\mathrm{bg}$} \\
   &  &  &  &  \\
  \hline
  $100$ & $1$ & $\agt 100 \%$ & $257$ & $3.6 \times 10^5$ \\
  & $10$ & $\agt 100 \%$ & & $3.6 \times 10^6$ \\
  \hline
  $300$ & $1$ & $\agt 100 \%$ & $770$ & $1.08 \times 10^6$ \\
  & $10$ & $\agt 100 \%$ & & $1.08 \times 10^7$ \\
  \hline
  $1000$ & $1$ & $\agt 100 \%$ & $2567$ & $3.6 \times 10^6$ \\
  & $10$ & $\agt 100 \%$ & & $3.6 \times 10^7$ \\
   \hline
  $10 \times100$ & $1$ & $\agt 100 \%$ & $2567$ & $3.6 \times 10^6$ \\
  & $10$ & $\agt 100 \%$ & & $3.6 \times 10^7$ \\
  \hline 
\end{tabular}} \\
 \vspace{0.4cm}
\footnotesize{
 \textbf{CTA \; E$_{\rm \bf th}$=100 GeV}  
 \vspace{0.1cm}
\begin{tabular}{|llrrrrc|}
  \hline
  \multirow{2}{*}{Observation time [h]} & \multirow{2}{*}{Bg. rate [Hz]} & 
  \multicolumn{2}{c}{${\bf \sigma_\mathrm{\textbf{fov}}=4^\circ}$} & \multicolumn{2}{c}{${\bf \sigma_\mathrm{\textbf{fov}}=5^\circ}$} &
  \multirow{2}{*}{$N_\mathrm{bg}$} \\ 
   &  &  Sens. & $N_\mathrm{sig}$ & Sens. & $N_\mathrm{sig}$ & \\
  \hline
  $100$ & $10$ & 90\% & 6687 & $70 \%$ & $10448$ & $3.6 \times 10^6$ \\
  & $100$ &\agt 100\%  &  & $\agt 100 \%$ & & $3.6 \times 10^7$ \\
  \hline
  $300$ & $10$ & 45\% & 20059  &  $35 \%$ & $31343$ & $1.08\times 10^7$ \\
  & $100$ & \agt 100\% &  &  $\agt 100 \%$ & & $1.08\times 10^8$ \\
  \hline
  $1000$ & $10$ & 30\% & 66867 & $20 \%$ & $104476$ & $3.6 \times 10^7$ \\
  & $100$ & 95\% &  & $ 75 \%$ & & $3.6 \times 10^8$ \\
   \hline
  $10 \times 100$ & $10$ &  55 \%  &66867  & $40 \%$ & $104476$ & $3.6 \times 10^7$ \\
  & $100$ & \agt 100\% &  & $\agt 100 \%$ & & $3.6 \times 10^8$ \\
  \hline 
\end{tabular}} \\
 \vspace{0.4cm}
\footnotesize{
 \textbf{CTA \; E$_{\rm \bf th}$=300 GeV}  
\begin{tabular}{|llrrrrc|}
  \hline
  \multirow{2}{*}{Observation time [h]} & \multirow{2}{*}{Bg. rate [Hz]} & 
  \multicolumn{2}{c}{${\bf \sigma_\mathrm{\textbf{fov}}=4^\circ}$} & \multicolumn{2}{c}{${\bf \sigma_\mathrm{\textbf{fov}}=5^\circ}$} &
  \multirow{2}{*}{$N_\mathrm{bg}$} \\ 
   &  &  Sens. & $N_\mathrm{sig}$ & Sens. & $N_\mathrm{sig}$ & \\
  \hline
  $100$ & $1$ & 55\% & 4262 & $30 \%$ & $6659$ & $3.6 \times 10^5$ \\
  & $10$ &\agt 100\% &  & $95 \%$ & & $3.6 \times 10^6$ \\
  \hline
  $300$ & $1$ & 30\% &  12785 & $20 \%$ & $19976$ & $1.08\times 10^6$ \\
  & $10$ & 80\%  &  & $60 \%$ & & $1.08\times 10^7$ \\
  \hline
  $1000$ & $1$ & 15\% & 42617 & $10 \%$ & $66587 $ & $3.6 \times 10^6$ \\
  & $10$ & 45\% &  & $30 \%$ & & $3.6 \times 10^7$ \\
   \hline
  $10 \times100$ & $1$ & 30\%  &  42617& $20 \%$ & $66587 $ & $3.6 \times 10^6$ \\
  & $10$ & 85\% &  & $65 \%$ & & $3.6 \times 10^7$ \\
  \hline 
\end{tabular}}
\end{center}
\caption{Sensitivity for detecting a self-annihilating DM contribution
  to the EDGB, utilizing anisotropy measurements in terms of the
  angular power spectrum. The sensitivity is given in terms of the
  minimum detectable DM gamma-ray flux, expressed in percentages of
  the EDGB. The three tables list the sensitivities for instrumental
  setups resembling current IACTs and the planned CTA, respectively,
  for several observation times, background rates, fov, and
  observational strategies. For reference, the number of simulated
  signal ($N_\mathrm{sig}$) and background ($N_\mathrm{bg}$) events is
  listed. A value of $\agt 100\%$ refers to a sensitivity outside our
  tested range.}
\label{tab_sens}
\end{table}

Finally, it should be noted that the sensitivity estimation described
above assumes previous knowledge of the intrinsic astrophysical and DM
anisotropies. In realistic cases, a measurement of these quantities
from the analysis itself would be preferable in order to perform a
comparison with the model predictions.  This can be done, in practice,
analyzing the data in multiple energy bands.  The presence of more
components at different energies and the intrinsic anisotropies of the
components can then be inferred from a study of the anisotropy energy
spectrum in intensity and fluctuation
\cite{SiegalGaskins:2009ux,Cuoco:2010jb}.  We do not pursue such
detailed analysis here, while for the sensitivity estimate to the DM
contribution the above approach is sufficiently accurate.

The results are shown in table \ref{tab_sens}. Clearly, current
instruments have limited prospects of constraining DM through
small-scale angular anisotropies. It is worth stressing, nonetheless,
that anisotropy searches also constrain population properties of
astrophysical sources (as discussed in more detail below), so that
this particular result should not hamper a dedicated search for
anisotropies with current observatories.  However, prospects for DM
searches improve for CTA, especially if the background rate can be
kept reasonably low. Sensitivities of $\sim20\%$ and $\sim10\%$ at
100\,GeV and 300\,GeV, respectively, can be achieved with
single-target observations of 1\,000\,h.  Even in the most
conservative case of a fov of 3$^\circ$, a sensitivity of $\sim30\%$
above 300 GeV can still be reached with CTA (not shown in the table),
while a fov of 4$^\circ$ would result in a sensitivity of $\sim15\%$.

\begin{figure}[t]
\centering
    \begin{subfigure}[t]{0.47\textwidth}
    \centering
    \includegraphics[width=\textwidth]{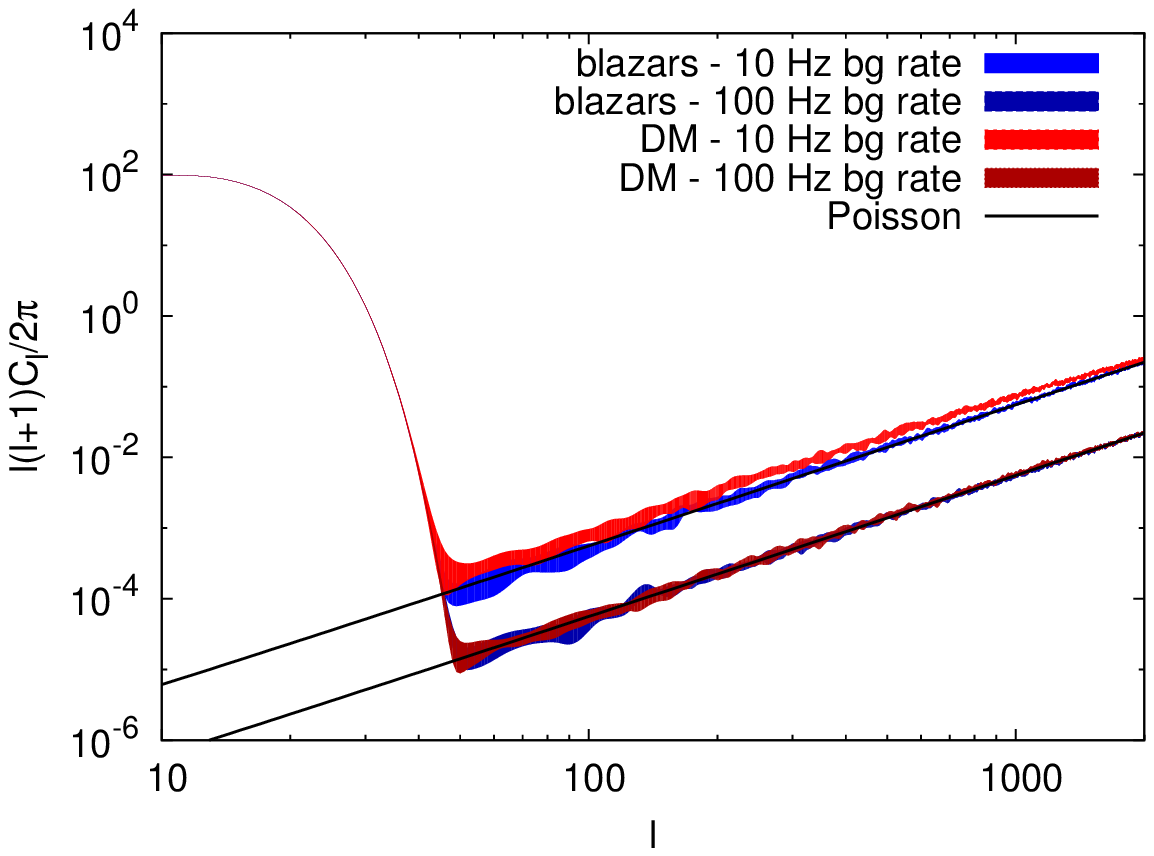}
    \caption{Single fov, $1\,000$\,h, $E_\mathrm{thr}=100\,\mathrm{GeV}$ \vspace{0.75cm}}
    \end{subfigure}
    \hspace{0.3cm}
    \begin{subfigure}[t]{0.47\textwidth}
    \centering
    \includegraphics[width=\textwidth]{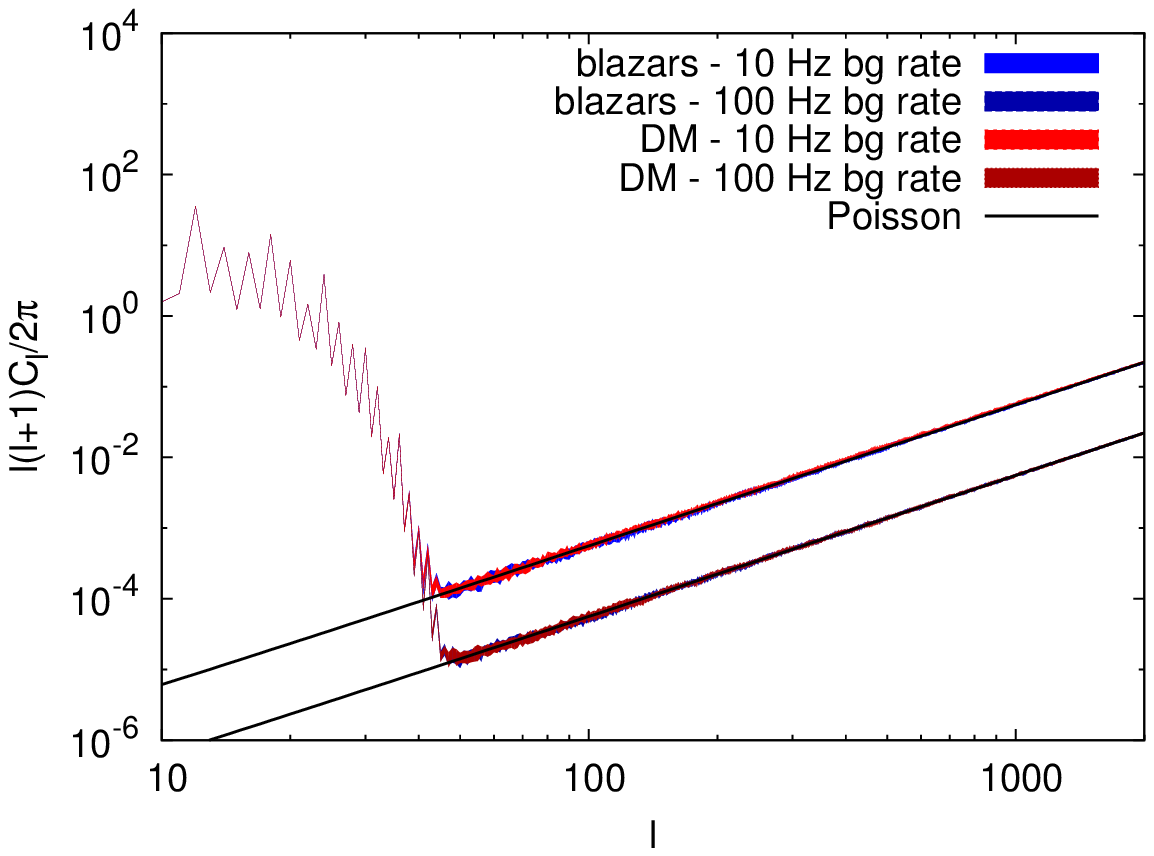}
    \caption{Multiple fov, $10\times 100$\,h, $E_\mathrm{thr}=100\,\mathrm{GeV}$}
    \end{subfigure}
    \begin{subfigure}[t]{0.47\textwidth}
    \centering
    \includegraphics[width=\textwidth]{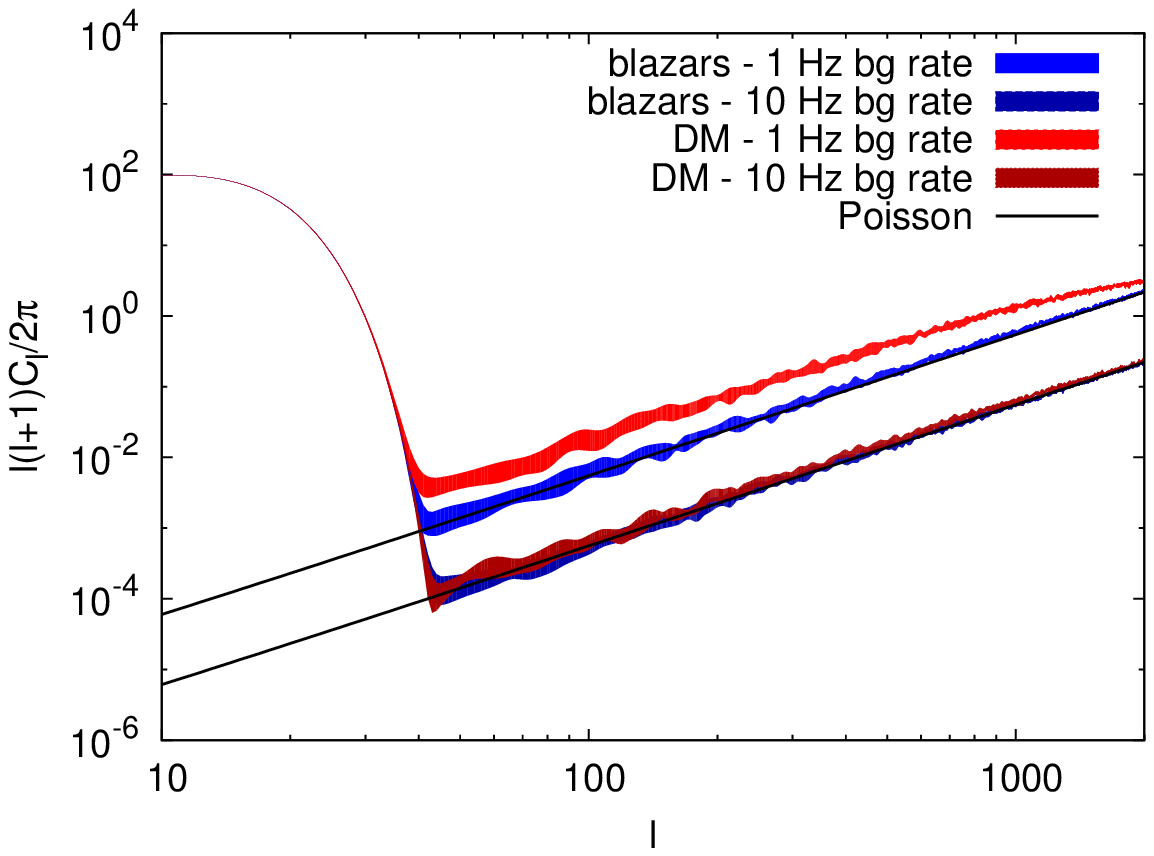}
    \caption{Single fov, $1\,000$\,h, $E_\mathrm{thr}=300\,\mathrm{GeV}$}
    \end{subfigure}
    \hspace{0.3cm}
    \begin{subfigure}[t]{0.47\textwidth}
    \centering
    \includegraphics[width=\textwidth]{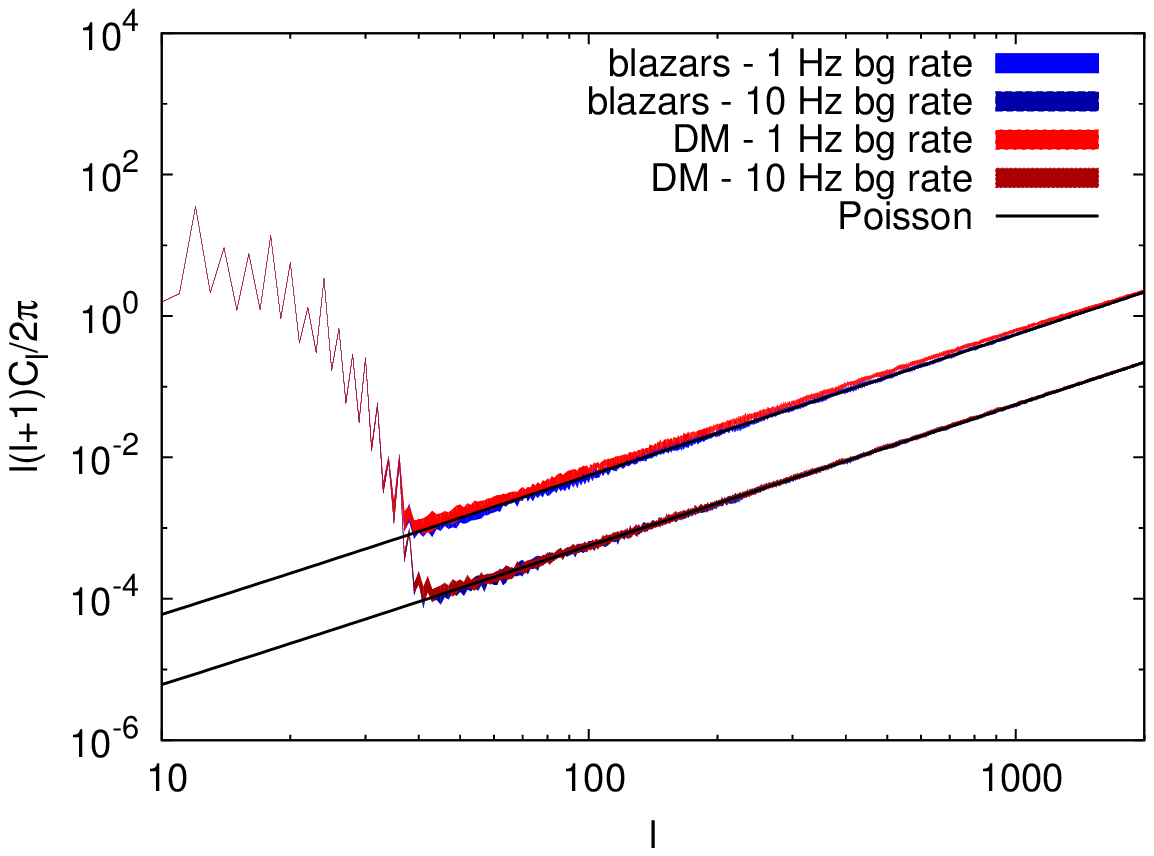}
    \caption{Multiple fov, $10\times 100$\,h, $E_\mathrm{thr}=300\,\mathrm{GeV}$}
    \end{subfigure}
\caption{Comparison between the measured APS for a pure astrophysical
  case with $C_{\ell} = C_P^{A}= 10^{-5}$ (blue bands) and a case with
  $40\%$ of the total radiation originating from self-annihilating DM
  with $C_{\ell} =C_P^{DM}= 10^{-3}$ (red bands).  An observation with
  a CTA-like telescope system of $1\,000 \, \text{h}$ on a single
  target (left column) and of $10 \times 100 \, \text{h}$ splitted on
  ten different targets (right column) is considered.  The upper plots
  refer to an energy threshold of 100\,GeV and the lower ones refer to
  a threshold of 300\,GeV. The two cases in each panel refer to
  background rates of $10 \, \text{Hz}$ and $100 \, \text{Hz}$ for
  100\,GeV, and $1 \, \text{Hz}$ and $10 \, \text{Hz}$ for
  300\,GeV. The size of the fov is $\sigma_\mathrm{fov} =
  5^\circ$. The lines show the estimated noise levels.}
\label{fig_reality1}
\end{figure}

\begin{figure}[t]
\centering
    \begin{subfigure}[t]{0.47\textwidth}
    \centering
    \includegraphics[width=\textwidth]{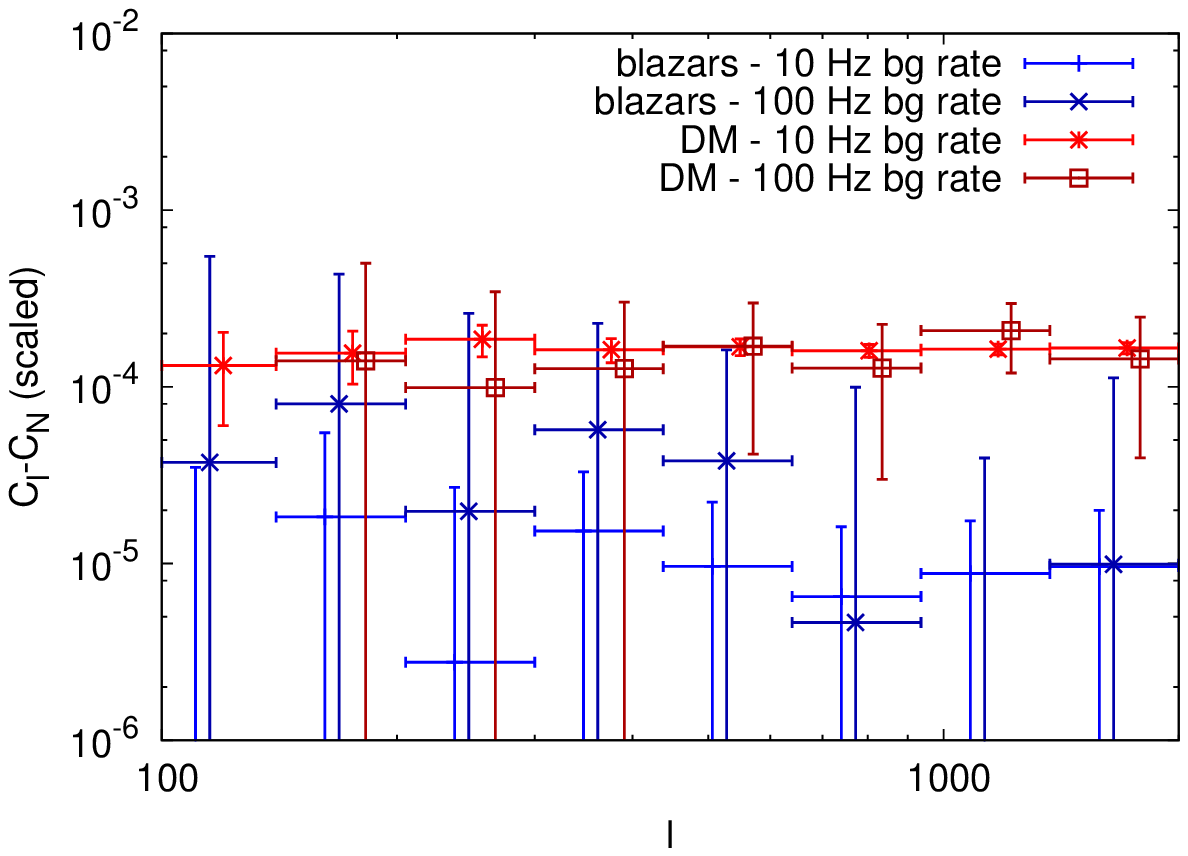}
    \caption{Single fov, $1\,000$\,h, $E_\mathrm{thr}=100\,\mathrm{GeV}$ \vspace{0.75cm}}
    \end{subfigure}
    \hspace{0.3cm}
    \begin{subfigure}[t]{0.47\textwidth}
    \centering
    \includegraphics[width=\textwidth]{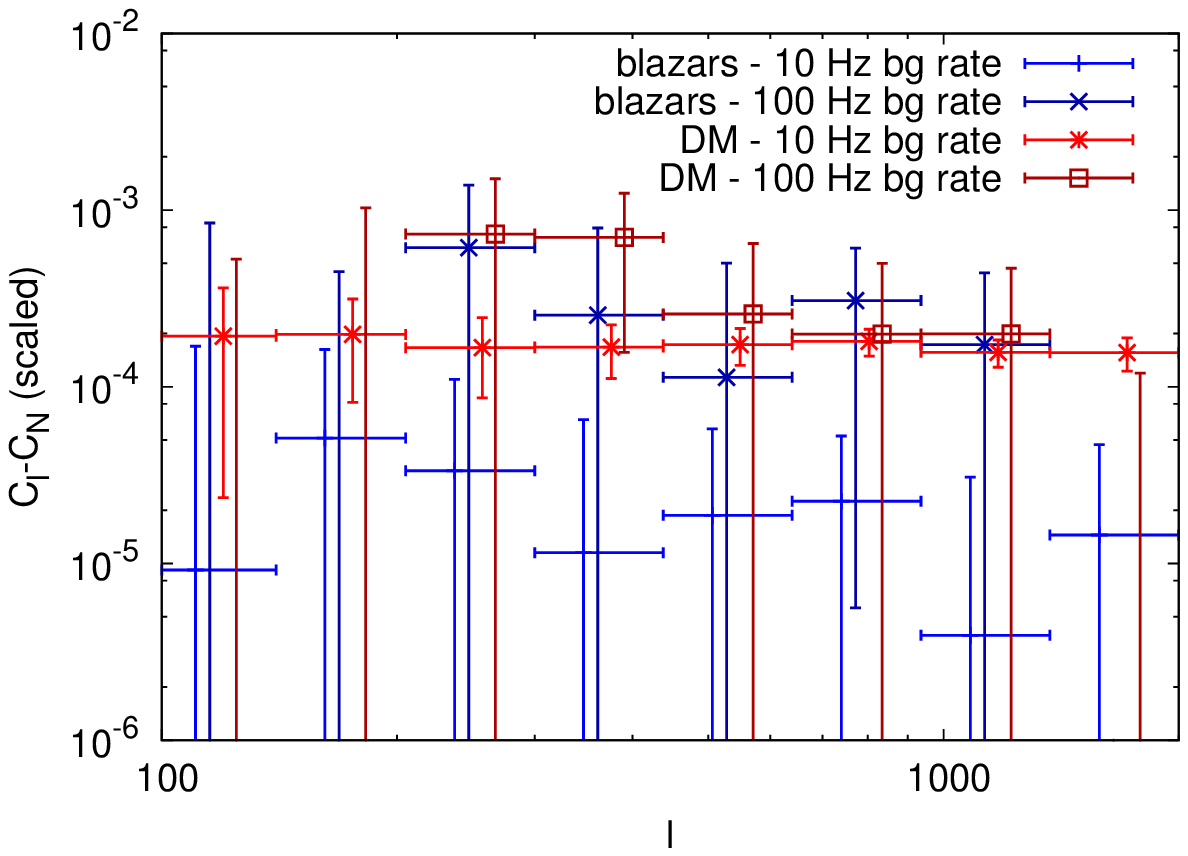}
    \caption{Multiple fov, $10\times 100$\,h, $E_\mathrm{thr}=100\,\mathrm{GeV}$}
    \end{subfigure}
    \begin{subfigure}[t]{0.47\textwidth}
    \centering
    \includegraphics[width=\textwidth]{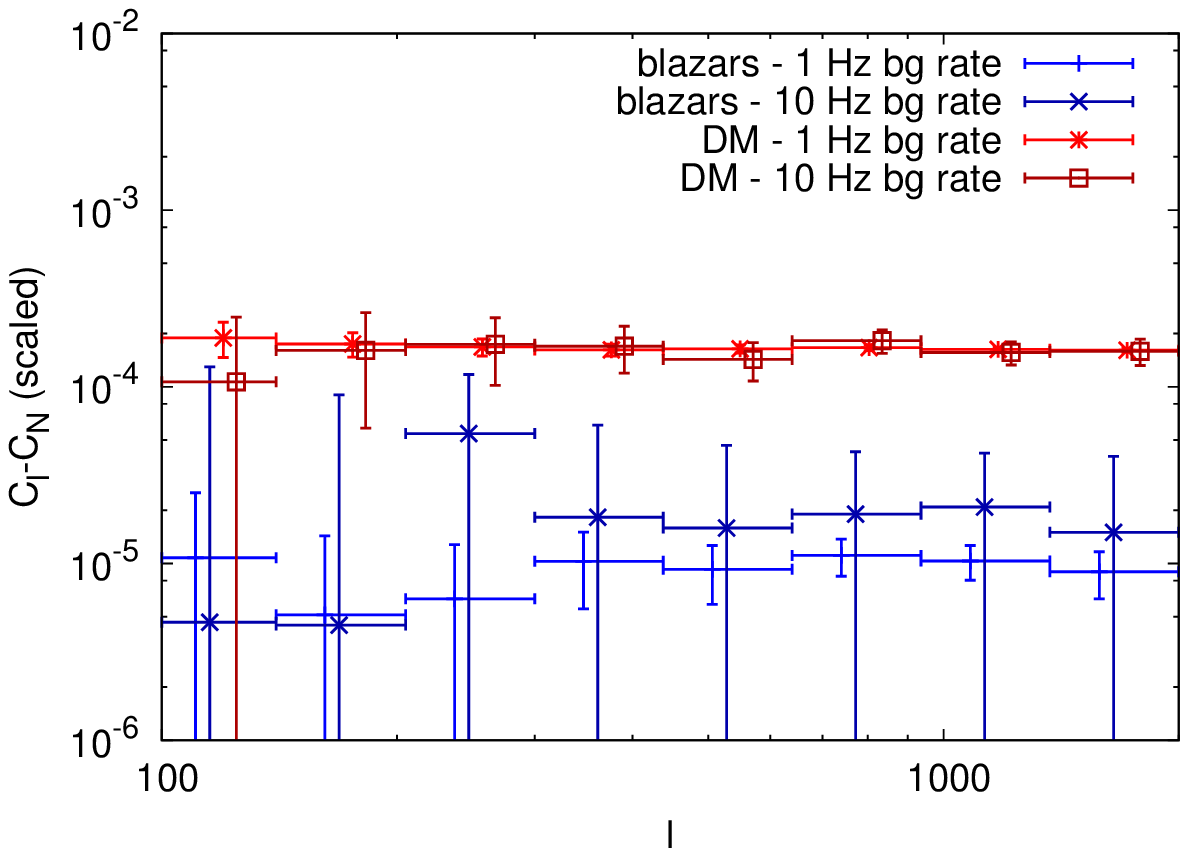}
    \caption{Single fov, $1\,000$\,h, $E_\mathrm{thr}=300\,\mathrm{GeV}$}
    \end{subfigure}
    \hspace{0.3cm}
    \begin{subfigure}[t]{0.47\textwidth}
    \centering
    \includegraphics[width=\textwidth]{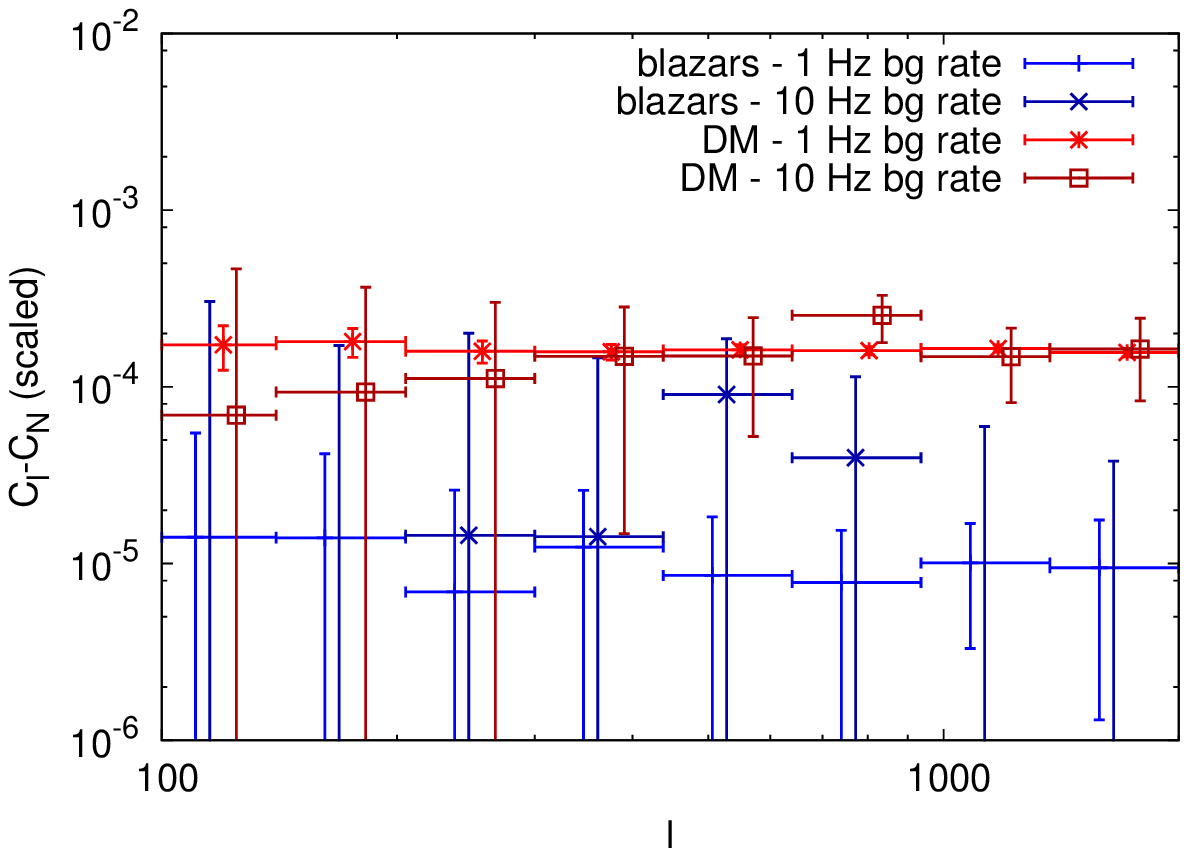}
    \caption{Multiple fov, $10\times 100$\,h, $E_\mathrm{thr}=300\,\mathrm{GeV}$}
    \end{subfigure}
\caption{Same as figure~\ref{fig_reality1}, but showing the
  noise-subtracted and instrumental effects unfolded APS. Note that,
  in difference to figure~\ref{fig_reality1}, the $y$-axis shows
  $C_{\ell} - C_N$ instead of $\ell(\ell+1)C_\ell/2\pi$. Also, the APS
  are binned into 8 logarithmically spaced bins in $\ell$. See text
  for more details.  For readability, the bins for each sub-case in
  each plot are slightly shifted.}
\label{fig_binnedAPS}
\end{figure}

Facing realistic data sets, the change in sensitivity  for different
observational strategies is worth mentioning.  As shown in
table~\ref{tab_sens}, the sensitivity for a combination of ten
different observations of 100\,h each is just a factor of 2 worse
compared to a continuous 1\,000\,h single target observation. In
addition, the sensitivity of the combined data set is approximately
equivalent to a full 300\,h observation of a single target.  Following
the discussion at the end of section~\ref{sec_res} we see that, even
with CTA, the statistical regime where a multiple-fov strategy results
in a reduction of the errors cannot be reached. Rather, the
combination of multiple fov results in errors comparable or slightly
larger than the single fov observation. Nonetheless, observations of
single targets for 1\,000\,h each are practically not feasible
(perhaps apart from the Galactic Center over several years), and the
observation of ten different targets for 100\,h each is more likely to
be realized. We emphasize that these observations do not need to
specifically target anisotropy searches, but observations taken for
different purposes can be analyzed instead.  In this manner, the loss
in sensitivity by a factor of $\sim\!2$ is compensated by the
availability of a significantly larger data set.

As a specific example of our analysis, figures~\ref{fig_reality1}
and~\ref{fig_binnedAPS} show the APS for the cases of 1\,000\,h and
10$\times$100\,h of observation time with CTA for the threshold
energies of 100\,GeV and 300\,GeV, and a fov of $5^\circ$. The pure
astrophysical case ($C_P^{A}= 10^{-5}$) and a case of a $40\%$ DM
($C_P^{DM}= 10^{-3}$) contribution are shown, as well as the two
different choices of background rates. Figure~\ref{fig_reality1} shows
the raw measured APS and illustrates the fact that the variance of the
APS decreases when the observation time is splitted over several fov,
but, at the same time, the signal over noise ratio also decreases.
Further informations are difficult to infer from
figure~\ref{fig_reality1}. For this reason unfolded and binned APS are
shown in figure~\ref{fig_binnedAPS}.  Unfolding involves subtraction
of the Poissonian noise and correction for the PSF attenuation. In the
simple case of a Gaussian beam, which we adopted, the PSF correction
is simply given by the factor $w_{\ell}^2= \exp (\sigma_{\rm psf}^2\,
\ell (\ell+1) )$, see \cite{Ackermann:2012uf,Ando:2005xg} and
appendix~\ref{app:a}. The presence of the background also alters the
normalization of the APS. This effect can be corrected by rescaling
the noise-subtracted APS by the factor $N_{\rm ev}^2/N_{\rm
  sig}^2$. Clearly, this correction requires a perfect, and thus
unrealistic, knowledge of the fraction of signal and background in the
collected data, while this information will likely be available with a
large error only. However, this will not pose a problem in the
analysis of real data, where the \emph{intensity} APS is used instead
of the dimensionless APS and no a priori knowledge of the gamma-ray
flux is required. Finally, the limited fov also introduces a
re-normalization of the APS by a factor $f_{\rm sky}$ which we
corrected in the unfolding. A binning of data is implemented given the
large error for a single multipole.  Since $C_\ell$ coefficients of
neighboring multipoles are correlated, the knowledge of the full
covariance matrix is in principle necessary to calculate the error
after binning.  Here, we use a simple analytical estimate of the
error, which is given by $\delta C_{\bar{\ell}} = (C_{\bar{\ell}}
+C_N\, w_{\bar{\ell}}^2) \sqrt{2/(2\bar{\ell}+1)/\Delta \ell/f_{\rm
    sky} }$, where $\Delta \ell$ is the width of the bin, $\bar{\ell}$
is the average $\ell$ of the bin, and $w_\ell=\exp ( \sigma_{\rm
  psf}^2\, \ell (\ell+1)/2 )$. This approach is accurate for all but
very low ($\ell<10$) multipoles. The unfolded APS of
figure~\ref{fig_binnedAPS} show that we recover within the errors the
input anisotropy $C_P^{A}= 10^{-5}$ for the astrophysical case and the
anisotropy $C_P^{\text{A}+\text{DM},40\%} = 0.4^2 \cdot C_P^{DM}
\!+0.6^2 \cdot C_P^{A} \simeq 1.6 \times 10^{-4}$ for the case of 40\%
DM. It can be also seen from the plot that, when the background rate
increases, the input signal can still be recovered although, as
expected, with a larger error.  The case of ten fov observations
explicitly shows that the errors worsen by approximately a factor of 2
in agreement with the results of table 2.  The sensitivity to DM
inferred from figure~\ref{fig_binnedAPS} is in line with the values
reported in table 2.

The other crucial parameters determining the sensitivity are
$C_P^{DM}$ and $C_P^A$.  To test the dependence on these parameters,
we performed a further simulation with $C_P^A=10^{-4}$ (instead of
$C_P^A=10^{-5}$), keeping the value $C_P^{DM}=10^{-3}$, and we found
the sensitivities to decrease by a factor $\sim\!3$. This seems to be
in good agreement with the analytic scaling relation
$(C_P^A/C_P^{DM})^{1/2}$ found in the appendix.  Given the strong
dependence on these two parameters, a firmer prediction of the
sensitivity requires pinning down their uncertainties. More accurate
calculations of $C_P^{DM}$ have been recently reported in
\cite{Fornasa:2012gu}, indicating that $C_P^{DM}$ can be as high as
$10^{-1}$, dominated by the contribution of the galactic substructures
over extragalactic ones (see in particular their figure~7).  Although
such a large DM anisotropy would push the sensitivity to values better
than $1\%$, the intrinsic emission from these very anisotropic
structures is expected to be very low, as witnessed by the fact that
the \emph{intensity} anisotropies are instead dominated by the
extragalactic component (see \cite{Fornasa:2012gu}).  The results
presented in \cite{Fornasa:2012gu} also indicate that the DM APS is
not exactly flat in multipole, but shows a slight attenuation to
higher multipoles. Given the good angular resolution of CTA, this
effect can in principle be used to disentangle the DM contribution
from the astrophysical one. However, the detection of a non-zero
anisotropy will likely be at a low signal-to-noise ratio, therefore it
will be difficult to explore large $\ell$ since they will be
noise-dominated. The effect is also somewhat degenerated with the PSF
attenuation and will thus require a good calibration of the
instrumental performance.  A more accurate estimate of $C_P^A$,
instead, awaits a direct measurement with Cherenkov telescopes or
further work on the modeling of blazar populations at TeV energies.
In this respect, the examples depicted in figures~\ref{fig_reality1}
and~\ref{fig_binnedAPS} show that the observation of an anisotropy
$C_P^A=10^{-5}$ is close to the sensitivity attainable with 1\,000\,h
of observation time with CTA (for the optimistic background estimate),
while it would be more challenging for the splitted observation
strategy.  However, $C_P^A$ is not precisely known, and the
possibility of a larger power such as $10^{-4}$ would obviously
improve its detection capabilities, albeit implying a lower DM
sensitivity.  We thus propose explicit observations of these kind of
anisotropies even with the current generation of instruments. Besides
the possibility of constraining a DM contribution, an observation of
anisotropy would provide a complementary and powerful tool to
investigate astrophysical TeV sources.

We checked that the results are robust with respect to the
approximation of a delta-like source flux distribution
$\mathrm{d}N/\mathrm{d}S$, as opposed to a more realistic
distribution, which typically shows a power law or a broken power-law
behavior (see for example \cite{Collaboration:2010gqa}). We simulated
the case of a power-law distribution $\propto S^{-2.4}$ for both DM
and astrophysical sources, spanning two orders of magnitude in flux
and normalized to give the same level of anisotropy as in the delta
case, i.e. $C_P^A=10^{-5}$ and $C_P^{DM}=10^{-3}$. The resulting
sensitivities are identical to the delta-like case. This result is
perhaps not surprising, since the anisotropy $C_P$ is given by an
integral over $\mathrm{d}N/\mathrm{d}S$ below the point-source
detection threshold $S_\mathrm{lim}$, more precisely $C_P=
\int_0^{S_\mathrm{lim}} \!\mathrm{d}S \, S^2 \mathrm{d}N/\mathrm{d}S$,
so that different $\mathrm{d}N/\mathrm{d}S$ distributions can still
result in the same $C_P$.

\begin{figure}[t]
\centering
    \begin{subfigure}[t]{0.47\textwidth}
    \centering
    \includegraphics[width=\textwidth]{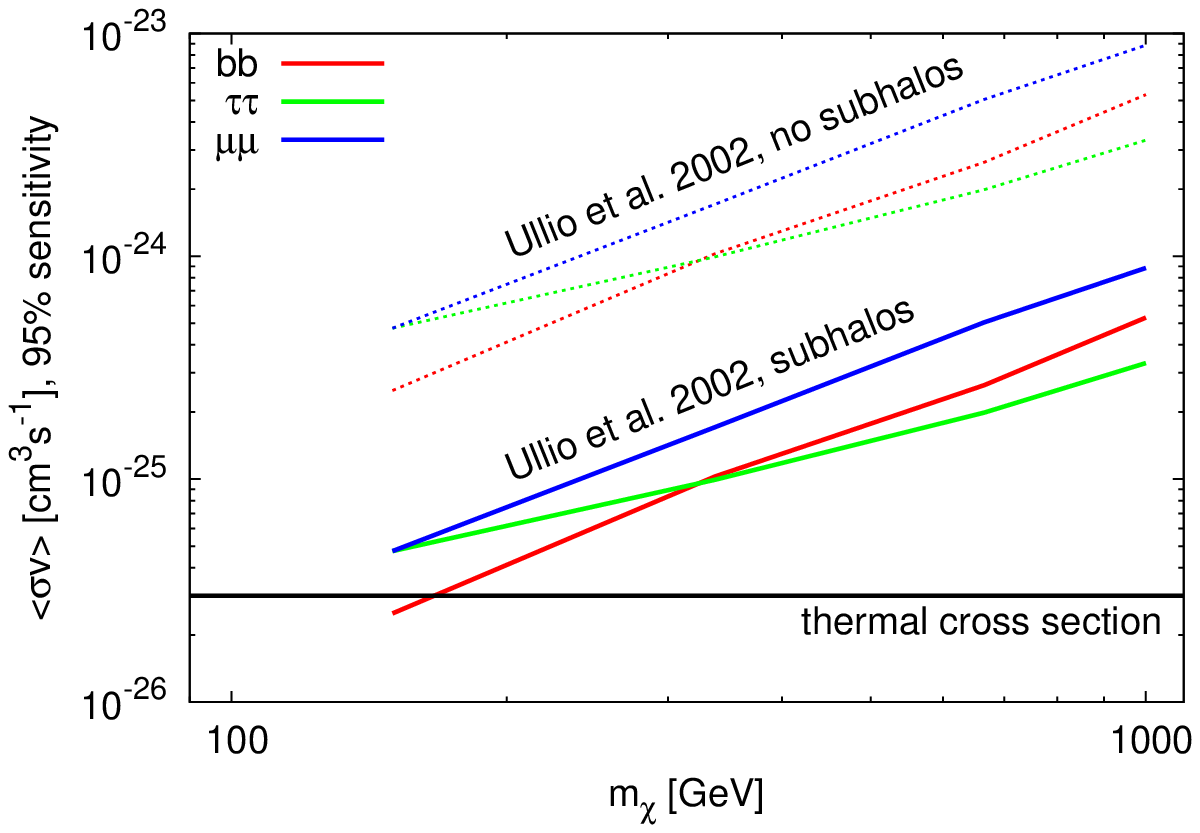}
    \caption{}
    \end{subfigure}
    \hspace{0.3cm}
    \begin{subfigure}[t]{0.47\textwidth}
    \centering
    \includegraphics[width=\textwidth]{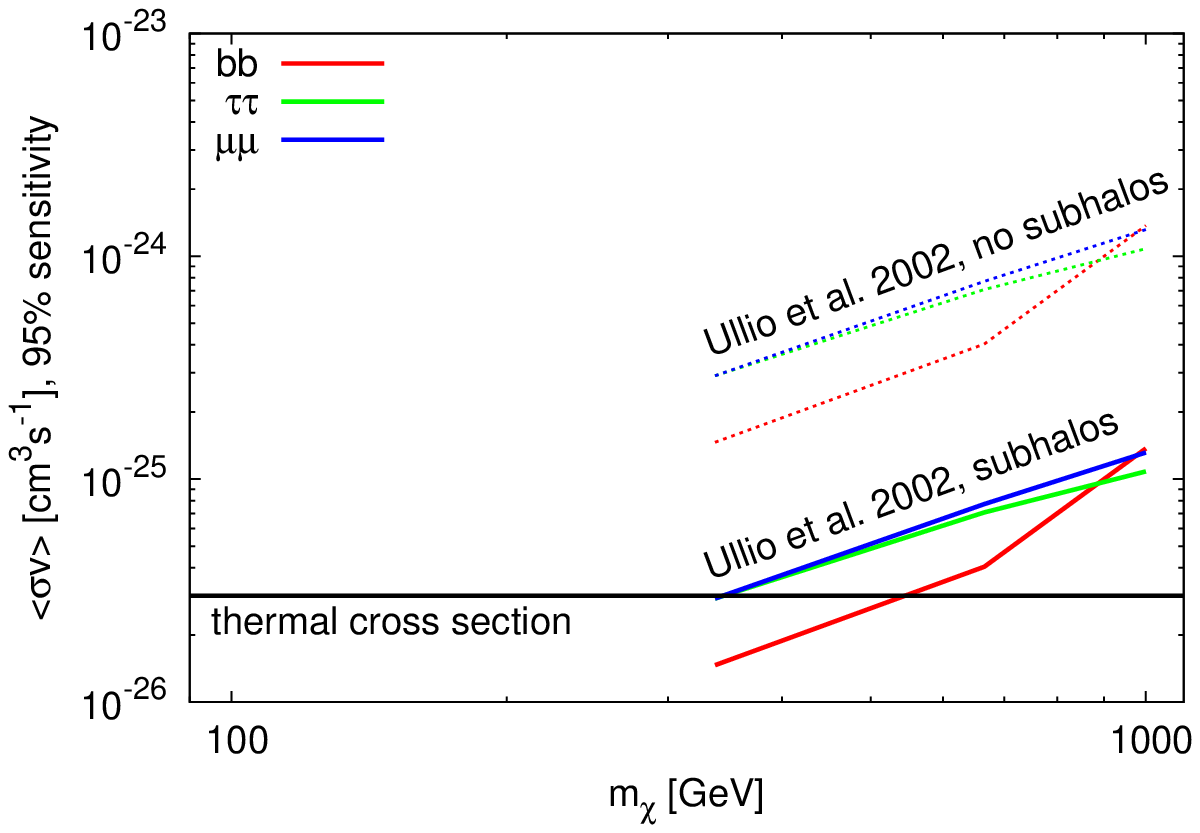}
    \caption{}
    \end{subfigure}
\caption{(a): sensitivity of CTA on the velocity-averaged DM
  self-annihilation cross section $\langle \sigma v \rangle$ as a
  function of the DM particle mass $m_\chi$, for an energy threshold
  of 100\,GeV, 300\,h (or $10\times 100$\,h) of observation time,
  $\sigma_\mathrm{fov}=5^\circ$, and a background rate of 10\,Hz
  (corresponding to a contribution from self-annihilating DM of
  $\sim\!35\%$ to the total EDGB, see table~\ref{tab_sens}). Solid
  lines correspond to the model of \cite{Ullio:2002pj},
  i.e. incorporating DM annihilation in subhalos, while dotted lines
  refer to the more conservative case of absent DM subhalos, giving a
  sensitivity by approximately a factor of 10 worse. We consider DM
  annihilation in $b\overline{b}$ (red), $\tau^+\tau^-$ (green), and
  $\mu^+\mu^-$ (blue) final states. Assumed instrumental
  characteristics are discussed in the text. (b): same as above for an
  energy threshold of 300\,GeV, 300\,h (or $10\times 100$\,h) of
  observation time, and a background rate of 1\,Hz (corresponding to a
  contribution from self-annihilating DM of $\sim\!20\%$ to the total
  EDGB, see table~\ref{tab_sens}).}
\label{fig_sens}
\end{figure}

The sensitivities given in table \ref{tab_sens} as fraction of the
EDGB flux can be expressed in terms of the more common quantity of the
velocity-averaged DM self-annihilation cross section $\langle \sigma v
\rangle$, although this introduces further model dependence. To
normalize the DM signal, we use the cosmological DM model of
\cite{Ullio:2002pj} in its optimistic version, where DM halos are
modeled with NFW profiles and include the presence of DM subhalos,
boosting the annihilation signal further (see, e.g.,
\cite{Zechlin:2011kk,Zechlin:2012by} for gamma-ray implications of
galactic DM subhalos). In addition, the less optimistic version
neglecting DM subhalos is considered, resulting in a factor of
$\sim\!10$ less DM annihilation flux and thus a correspondingly worse
sensitivity. It should be noted that considering the results of
\cite{Zavala:2009zr} from the Millennium-II simulation, an order of
magnitude enhancement with respect to the above ``optimistic'' case is
predicted with a correspondingly improved sensitivity (see also
\cite{Abdo:2010dk}, in particular figure~1). The results are shown in
figure~\ref{fig_sens} for various annihilation channels as function of
the DM particle mass $m_\chi$. In conclusion, the self-annihilation
cross section $\langle \sigma v \rangle = 3 \times 10^{-26} \,
\text{cm}^{3} \text{s}^{-1}$ expected from thermal dark matter
freeze-out can be probed with CTA for DM particle masses up to a few
hundred GeV. Interestingly, even using the conservative version of the
EDGB DM model above, sensitivities better than the ones achievable
with dwarf spheroidal galaxies observed with CTA and comparable with
achievable limits using cluster observations can be reached
\cite{Doro:2012xx}.

\section{Conclusions} \label{sec_dis}
We have investigated the key aspects of the capability of ground-based
gamma-ray telescopes with small fields of view (i.e. imaging
atmospheric Cherenkov telescopes) for observing anisotropies in the
diffuse gamma-ray background. In particular, the effects of the
effective area, the field of view, the angular resolution (PSF), and
of the hadron-background rejection efficiency have been studied. These
properties have been identified as crucial instrumental
characteristics, determining the sensitivity for detecting small-scale
angular anisotropies.

The sensitivity for detecting a contribution of self-annihilating dark
matter to the diffuse gamma-ray background has been investigated,
focussing on the analysis of angular anisotropies. Benchmark
instrumental setups of currently operating imaging atmospheric
Cherenkov telescopes such as H.E.S.S., MAGIC, and VERITAS, as well as
the forthcoming CTA have been considered. We have used realistic
expectations for the anisotropy power spectra from self-annihilating
dark matter and from astrophysical sources. We find that the
sensitivity of CTA will be sufficient to resolve a relative
contribution of $\sim$10\% from self-annihilating dark matter to the
total isotropic gamma-ray background flux, given an observation time
of 1\,000\,h and a background rate of 1\,Hz above 300\,GeV. More
important, we find that with a multiple field-of-view strategy of
1\,000\,h of observation time splitted over ten separate targets of
100\,h each yields a slight reduction in sensitivity (to $\sim\!20\%$,
a factor of 2) only.  In practice, this means that it will be possible
to obtain interesting constraints on dark matter without dedicated deep
observations, but combining existing observations of different primary
astrophysics targets.  The sensitivity achievable can be already
sufficient to probe the thermal annihilation cross section for WIMP
masses $\alt 200$\,GeV (for common models of dark matter annihilation
in galactic and extragalactic environments). We also find that CTA
will have sufficient sensitivity for detecting small-scale
anisotropies from astrophysical sources with a Poissonian anisotropy
level of $10^{-5}$, while the sensitivity of current-generation
instruments is approximately an order of magnitude lower. Given the
uncertainty on the exact expected anisotropy level, we propose that
available deep exposures, preferably at high galactic latitudes,
should be analyzed in order to search for anisotropies.  An
observation of anisotropy would provide a complementary and invaluable
tool for investigating the nature of TeV sources.

\section*{Acknowledgments}
We wish to thank Mattia Fornasa and Jesus Zavala for providing
suggestions on the manuscript and for useful discussions. We kindly
thank the anonymous referee for useful and important suggestions on
improving the manuscript. DH acknowledges support through the
collaborative research center (SFB) 676 ``Particles, Strings, and the
Early Universe'' at the University of Hamburg. JC is Royal Academy of
Science Fellow financed by a grant of the Knut and Alice Wallenberg
foundation.

\appendix

\section{Error on the Poisson anisotropy}\label{app:a}
The error on the \emph{fluctuation} angular power spectrum is given by 
\begin{equation}
 \delta C_\ell \equiv \sigma_\ell = (C_\ell+C_N w_\ell^2)
 \sqrt{\frac{2}{f_\mathrm{sky}(2\ell+1)}}\,,
\end{equation}
where $C_N= \Omega_\mathrm{fov}(1/N_\gamma+N_b/N_\gamma^2)$ is the
(Poissonian) noise, $N_{\gamma}$ the number of gamma-ray events, $N_b$
the number of background events, and $\Omega_\mathrm{fov}$ the total
fov in steradians \cite{Ando:2005xg}; $w_\ell=\exp
(\sigma_\text{psf}^2 \ \ell (\ell+1)/2)$ describes the correction for a
Gaussian PSF of width $\sigma_\text{psf}$ (in radians).  In the
following derivation of the sensitivity, we assume a Poisson-like power
spectrum, i.e. $C_\ell=C_P$.  In practice, $C_P$ is estimated by
calculating the weighted average of the measured angular power
spectrum, so that $C_P= \sum_\ell p_\ell C_\ell$, where $p_\ell=
(1/\sigma_\ell^2)/(\sum_\ell 1/\sigma_\ell^2)$ (the weight of higher
multipoles is larger, owing to smaller errors until the PSF error
starts to dominate).  The corresponding error on $C_P$ is the given by
$(\delta C_P)^2= \sum_\ell p_\ell^2 (\delta C_\ell)^2 = 1/(\sum_\ell
1/\sigma_\ell^2)$.

\subsection{Simple calculation}
In the following, we consider both a sufficiently narrow PSF,
resulting in a sufficiently small PSF correction for the multipole
range of interest, and the high statistics limit, so that $C_N$ is
negligible with respect to $C_P$.  In this limit, we have (implicitly,
$\ell \gg 1$ is assumed)
\begin{equation}
\sigma_\ell = C_\ell \sqrt{\frac{1}{f_\mathrm{sky} \ \ell}}\,.
\end{equation}
Evaluating the sum in $(\delta C_P)^2= 1/(\sum_\ell 1/\sigma_\ell^2)$
(this can be done analytically or approximating the sum as an
integral) yields
\begin{equation}
 \delta C_P \approx C_P
 \sqrt{\frac{2}{f_\mathrm{sky}(\ell_\mathrm{max}^2-\ell_\mathrm{min}^2)}} \approx
 \frac{C_P}{\ell_\mathrm{max}} \sqrt{\frac{2}{f_\mathrm{sky}}}\,.
\end{equation}

\subsection{More accurate calculation}
Assuming $\ell\gg1$,
\begin{equation}
 \sigma_\ell =  (C_\ell+C_N w_\ell^2) \sqrt{\frac{1}{f_\mathrm{sky} \ \ell}}\,.
\end{equation}
Evaluating the sum $(\delta C_P)^2= 1/(\sum_\ell 1/\sigma_\ell^2)$ by
approximating it as an integral we find
\begin{equation}
 \delta C_P \approx \sqrt{\frac{1}{f_\mathrm{sky}}} \left(
 -\frac{\ell_\mathrm{min}^2}{C_P^2} -\frac{\ln\left(C_P/C_N +
   w_{\ell_\mathrm{min}}^2\right)}{C_P^2 \ \sigma_\mathrm{psf}^2}
 -\frac{1}{2\,C_P\,\sigma_\mathrm{psf}^2 \left( C_P + C_N w_{\ell_\mathrm{min}}^2
   \right) } \right)^{-\frac{1}{2}}\,.
\end{equation}
With $\ell_\mathrm{min}\approx 100$, $\sigma_\mathrm{psf}=0.05^\circ
\ll 1/\ell_\mathrm{min}\approx 0.6^\circ$, and in the limit of high
statistics ($C_P\gg C_N$), the expression can be simplified as
\begin{equation}
 \delta C_P \approx C_P \, \sigma_\mathrm{psf}
 \, \sqrt{\frac{2}{f_\mathrm{sky} \, \ln\left( \frac{C_P}{e\, C_N}
     \right) }} \equiv \frac{C_P}{\ell_\mathrm{max}}
 \sqrt{\frac{2}{f_\mathrm{sky}}}\,,
\end{equation}
which is equivalent to the previous simplified calculation with the
definition $\ell_\mathrm{max}= \sigma_\mathrm{psf}^{-1}\sqrt{\ln\left(
  \frac{C_P}{e\, C_N} \right)}$.  Note that the quantity $\ln\left(
\frac{C_P}{e\, C_N} \right)$ is of order $\sim\!1$ in the high
statistics limit. In the case discussed in the paper,
$\sigma_\mathrm{psf}= 0.05^\circ$ is assumed, so that the first
condition is satisfied.  Further, a fov of $5^\circ$ corresponds to
$\Omega_\mathrm{fov}\approx10^{-2}$\,sr and
$f_\mathrm{sky}\approx10^{-3}$. Given a 1\,000\,h observation and a
background rate of 10\,Hz above 100\,GeV, $C_N\approx 3
\times10^{-5}$, referring to the number of events in
table~\ref{tab_sens}, while the case for 1\,Hz above 300\,GeV
corresponds to $C_N\approx 8 \times10^{-6}$, both to be compared with
$C_P=10^{-5}$. Therefore, we face the condition $C_P \sim C_N$ even
for the most optimistic cases.  Nonetheless, for illustration
purposes, the following derivation focusses on the regime $C_P\gg
C_N$, estimating the sensitivity achievable under optimal conditions.

\section{Sensitivity using the intensity APS}
More convenient calculations can be carried out in terms of the
intensity APS, since power spectra are linearly additive in case of
uncorrelated summands.  This is indeed the case for Poisson-like
anisotropies.

The intensity APS is simply related to the fluctuation APS as
\begin{equation}
 C_P^I=I^2\,C_P,
\end{equation}
where $I$ is the intensity of the considered component. We consider
the same scenario as above, i.e. an astrophysical component with
Poissonian anisotropy $C_A$ and intensity $I_A$, and a DM component
with anisotropy $C_{DM}$ and intensity $I_{DM}$. The corresponding
intensity anisotropies are given by $C_A^I=I_A^2\,C_A$ and
$C_{DM}^I=I_{DM}^2\,C_{DM}$. Assuming that the intensity anisotropy
$C_P^{I,\rm data}\pm \delta C_P^{I,\rm data}$ has been measured in a
given energy band, we can set a conservative upper limit on the DM
contribution as
\begin{equation}
 C^I_{DM}\alt C_P^{I, \rm data}\,,
\end{equation}
which implies
\begin{equation}
\frac{I_{DM}}{I} \alt \sqrt{\frac{C_P^{\rm data}}{C_{DM}}}\,.
\end{equation}
If the DM component is subdominant, we have $C_P^{\rm data} \sim C_A$
and $I_{DM}/I \alt \sqrt{C_A/C_{DM}}$. For the benchmark case with
$C_A=10^{-5}$ and $C_{DM}=10^{-3}$, this yields a maximum sensitivity
of $I_{DM}/I \alt 10\%$.

A case more accurately approximating the case discussed with the
previous simulations is given when the quantity $C_P^{I,\rm data} =
C_A^I$ is known within a certain error in advance, either because we
have measured this quantity in a different energy band, or because it
matches our theoretical expectations. In this case, a more interesting
upper limit can be derived from
\begin{equation}
 C^I_{DM}\alt \delta C_P^{I, \rm data}\,,
\end{equation}
which implies
\begin{equation}
\frac{I_{DM}}{I} \alt \sqrt{\frac{C_P^{\rm data}}{C_{DM}}} \times
\sqrt[4]{{\frac{2}{f_\mathrm{sky} \ \ell_\mathrm{max}^2}}}\,.
\end{equation}
Again, with $C_P^{\rm data}=C_A$ and for our benchmark case
$C_A=10^{-5}$, $C_{DM}=10^{-3}$, $\ell_\mathrm{max}\approx1000$, and
$f_\mathrm{sky}\approx10^{-3}$, we have a maximum sensitivity of
$I_{DM}/I \alt 2\%$. Note that the sensitivity improves with a larger
fov, although only with the fourth root. However, the sensitivity
improves faster with the angular resolution ($\ell_\mathrm{max}$).

We emphasize that the above results are clearly back-of-the-envelope
calculations, with the purpose of estimating the sensitivity and its
dependence on the relevant quantities. Dealing with real data, more
appropriate derivations can be conducted, for example, with a
likelihood analysis.

\section{Sensitivity using the fluctuation APS}
In the paper, fluctuation anisotropies are used to calculate the
sensitivities. In this case, the relation effectively imposed to
derive the sensitivity is
\begin{equation}
C_P^{\rm data} -\delta C_P^{\rm data} \alt C_\mathrm{tot} \alt C_P^{\rm data} +\delta C_P^{\rm data}\,,
\end{equation}
where
\begin{equation}
C_\mathrm{tot} = \frac{C_{DM} I_{DM}^2 + C_A I_A^2}{\left(I_{DM}+I_A\right)^2}\,.     
\end{equation}
Rewriting $C_P^{\rm data} \pm \delta C_P^{\rm data}$ as $C_P^{\rm
  data} \left(1 \pm \Delta \right)$, where $\Delta \equiv
\sqrt{2/(f_\mathrm{sky} \ell_\mathrm{max}^2)}$, we find that the
sensitivity is given by a second order equation, solved by
\begin{equation}\label{Eq_quadratic}
\frac{I_{DM}}{I_A} = \frac{C_P^{\rm data} \left(1 \pm \Delta \right)
  \pm \sqrt{C_P^{\rm data} \left(1 \pm \Delta \right) \left( C_A
    +C_{DM}\right) - C_A C_{DM}}}{C_P^{\rm data} \left(1 \pm \Delta
  \right) - C_{DM}}\,.
\end{equation}
The equation can be simplified in a few relevant cases. For example,
if $C_{DM} \gg C_A$ (as in the benchmark case), and again assuming
$C_P^{\rm data} \approx C_A$, the positive (physical) solution is
\begin{equation}
\frac{I_{DM}}{I} \alt \frac{C_P^{\rm data}}{C_{DM}} +
\sqrt{\frac{C_P^{\rm data}}{C_{DM}}}
\ \sqrt[4]{{\frac{2}{f_\mathrm{sky} \ell_\mathrm{max}^2}}}\,.
\end{equation}
For the benchmark numbers, we get a sensitivity of $I_{DM}/I \alt
3\%$, only slightly worse than in the intensity case. Therefore, the
actual value of 10\% for the sensitivity quoted in the paper for the
most favorable scenario is not very far from the analytical estimate.

For $C_A=10^{-4}$ and $C_{DM}=10^{-3}$ the above sensitivity degrades
quite rapidly, worsening by a factor of 10, in agreement with the
simulations performed.  In the intensity case, instead, the dependence
only goes with the square root, and the sensitivity should worsen by a
factor of $\sim$3 only.

Finally, unlike in the intensity case, we get an upper limit on
$I_{DM}$, even in the case $C_{DM} \ll C_A$. This scenario is quite
unphysical, although it can be practically taken as case study for a
further (non-DM) component with negligible anisotropy. This is
expected, for example, from normal galaxies or truly diffuse processes
like photons from UHECRs cascades on the CMB.  In this case, after a
few simplifications we get
\begin{equation}
\frac{I_{DM}}{I} \alt \sqrt{\frac{2}{f_\mathrm{sky} \ell_\mathrm{max}^2}}\,,
\end{equation}
which is independent of $C_A$. The benchmark numbers for
$f_\mathrm{sky}$ and $\ell_\mathrm{max}$ yield a limit of $I_{DM}/I
\alt 5\%$.

The case $C_A \sim C_{DM}$ seems to give the worst limit. For this
regime, a simple formula cannot be derived and the full expression
Eq.~\ref{Eq_quadratic} needs to be used.

\end{document}